\documentclass[final,authoryear,3p,times]{elsarticle}

\usepackage{graphicx,color,colortbl}
\usepackage{enumerate}
\usepackage[colorlinks]{hyperref}
\usepackage{amsmath,amssymb,amsthm,bm}
\usepackage{multicol}
\usepackage[dvipsnames]{xcolor}
\usepackage{ulem}
\usepackage{tikz}
\usepackage{svg}
\usetikzlibrary{automata,positioning}
\usepackage{afterpage}
\usepackage{calc}
\usepackage{ifthen}
\usepackage{algorithm}
\usepackage{algpseudocode}
\usepackage{caption}
\usepackage{subcaption}
\usepackage{siunitx}
\usepackage{booktabs}

\edef\svtheparindent{\the\parindent}
\usepackage{parskip}
\parindent=\svtheparindent\relax

\definecolor{ForestGreen}{RGB}{34,139,34}
\definecolor{InternationalOrange}{rgb}{1.0, 0.31, 0.0}

\newcommand{\CA}{}
\newcommand{\CB}{}
\newcommand{\CC}{}
\newcommand{\CD}{}

\newcommand{\RomanNumeralCaps}[1]
    {\MakeUppercase{\romannumeral #1}}

\newcommand{\eqname}{Equation~}
\newcommand{\eqnames}{Equations~}
\renewcommand{\figurename}{Figure~}
\renewcommand{\tablename}{Table~}
\newcommand{\sectionname}{Section~}

\newcommand{\idxLOAD}{t}

\newcommand{\idxREAC}{\beta}
\newcommand{\idxPHASE}{k}

\newcommand{\bfeps}{\boldsymbol{\varepsilon}}

\newcommand{\boldface}[1]{\mathbf{#1}}   

\newcommand{\bft}{\boldface{t}}
\newcommand{\bfu}{\boldface{u}}
\newcommand{\bfv}{\boldface{v}}

\newcommand{\bfF}{\boldface{F}}

\newcommand{\bfI}{\boldface{I}}

\newcommand{\bfX}{\boldface{X}}

\newcommand{\bfalpha}{\boldsymbol{\alpha}}

\newcommand{\bfepsilon}{\boldsymbol{\varepsilon}}

\newcommand{\bftheta}{\boldsymbol{\theta}}

\newcommand{\bfsigma}{\boldsymbol{\sigma}}

\newcommand{\calD}{\mathcal{D}}

\newcommand{\calN}{\mathcal{N}}

\newlength{\boxwidth}
\def\dd{\;\!\mathrm{d}}

\def\btheorem{\begin{theorem}}
\def\etheorem{\end{theorem}}
\def\blemma{\begin{lemma}}
\def\elemma{\end{lemma}}
\def\bproposition{\begin{proposition}}
\def\eproposition{\end{proposition}}
\def\bcorollary{\begin{corollary}}
\def\ecorollary{\end{corollary}}
\def\bdefinition{\begin{definition}}
\def\edefinition{\end{definition}}
\def\bexample{\begin{example}}
\def\eexample{\end{example}}
\def\bremark{\begin{remark}}
\def\eremark{\end{remark}}

\DeclareMathOperator{\tr}{tr}

\DeclareMathOperator{\dev}{dev}
\DeclareMathOperator{\vol}{vol}

\newcommand{\be}{\begin{equation}}
\newcommand{\ee}{\end{equation}}
\newcommand{\beq}{\begin{eqnarray}}
\newcommand{\eeq}{\end{eqnarray}}
\newcommand{\bem}{\begin{multline}}
\newcommand{\eem}{\end{multline}}
\newcommand{\ba}{\begin{align}}
\newcommand{\ea}{\end{align}}

\newcommand{\argminWithArgs}{\operatornamewithlimits{arg\ min}}

\begin{document}

\begin{frontmatter}

\title{Automated discovery of generalized standard material models with EUCLID}

\author[eth]{Moritz Flaschel}
\author[tud]{Siddhant Kumar}
\author[eth]{Laura De Lorenzis\corref{cor1}}

\cortext[cor1]{Correspondence: ldelorenzis@ethz.ch}

\address[eth]{Department of Mechanical and Process Engineering, ETH Z\"{u}rich, 8092 Z\"{u}rich, Switzerland}
\address[tud]{Department of Materials Science and Engineering, Delft University of Technology, 2628 CD Delft, The Netherlands}

\begin{abstract}
We extend the scope of our recently developed approach for unsupervised automated discovery of material laws (denoted as EUCLID) to the general case of a material belonging to an unknown class of constitutive behavior. To this end, we leverage the theory of generalized standard materials, which encompasses a plethora of important constitutive classes including elasticity, viscosity, plasticity and arbitrary combinations thereof. We show that, based only on full-field kinematic measurements and net reaction forces, EUCLID is able to automatically discover the two scalar thermodynamic potentials, namely, the Helmholtz free energy and the dissipation potential, which completely define the behavior of generalized standard materials. The a priori enforced constraint of convexity on these potentials guarantees by construction stability and thermodynamic consistency of the discovered model; balance of linear momentum acts as a fundamental constraint to replace the availability of stress-strain labeled pairs; sparsity promoting regularization enables the automatic selection of a small subset from a possibly large number of candidate model features and thus leads to a parsimonious, i.e., simple and interpretable, model. Importantly, since model features go hand in hand with the correspondingly active internal variables, sparse regression automatically induces a parsimonious selection of the few internal variables needed for an accurate but simple description of the material behavior. A fully automatic procedure leads to the selection of the hyperparameter controlling the weight of the sparsity promoting regularization term, in order to strike a user-defined balance between model accuracy and simplicity. By testing the method on synthetic data including artificial noise, we demonstrate that EUCLID is able to automatically discover the true hidden material model from a large catalog of constitutive classes, including elasticity, viscoelasticity, elastoplasticity, viscoplasticity, isotropic and kinematic hardening.

\end{abstract}


\end{frontmatter}



\section{Introduction}\label{sec:Introduction}
The conventional strategy to mathematically describe the response of a given material in solid mechanics entails the a priori choice of a material model depending on a finite number of tunable parameters, and the calibration of these parameters based on experimental observations. The calibrated material model, which describes the relation between stresses and strains (and potentially other material state variables), can then be deployed in numerical (e.g. finite element) simulations to predict the mechanical response of arbitrarily shaped components of the material under external influences. The weakest aspect of this procedure lies in the a priori selection of a suitable model, which is largely based on experience. An inappropriate initial choice inevitably leads to failure to identify a set of material parameters which enable the model to accurately represent the experimental data; in such case the initial choice has to be modified and the procedure repeated, possibly multiple times, resulting in a time-consuming and error-prone iterative process. 

Enabled by the tremendous amount of data made available by the recent advances in experimental mechanics, data-driven approaches, often based on machine learning tools, are increasingly being explored for applications in solid mechanics, and especially in material modeling (\cite{neggers_big_2018,montans_data-driven_2019,liu_machine_2021,peng_multiscale_2021,aldakheel_what_2022}).
A popular strategy to mitigate the drawbacks of an inappropriate a priori model selection is to choose a parametric material model ansatz that is as general as possible, i.e., with a large number of tunable parameters. Examples are splines \citep{sussman_model_2009}, Gaussian processes \citep{frankel_tensor_2020}, neural networks \citep{ghaboussi_knowledgebased_1991} or neural ordinary differential equations \citep{tac_data-driven_2022}. Due to the large number of trainable parameters and functions that describe the material response, these types of model ansatz are more flexible than conventional models and the risk of not being able to well interpret the observed data is reduced. Combinations of traditional and machine learning components can further increase the performance of the resulting material models, as shown for example by \cite{gonzalez_learning_2019,ibanez_hybrid_2019,fuhg_model-data-driven_2021,fuhg_elasto-plasticity_2022}. Recognizing that constitutive models should not violate any well-known physical requirements motivated the recent trend to supplement machine learning models with physics-based constraints. These are enforced either by construction or in a weak sense by introducing additional regularization terms to the loss function that penalize the violation of physical laws. Examples of constraints include convexity or polyconvexity in hyperelasticity \citep{asad_mechanics-informed_2022,klein_polyconvex_2022,klein_finite_2022,kalina_automated_2022,tac_data-driven_2022,thakolkaran_nn-euclid_2022} or the second law of thermodynamics for dissipative materials \citep{masi_thermodynamics-based_2021,vlassis_sobolev_2021,fuhg_elasto-plasticity_2022,huang_variational_2022,masi_evolution_2022}. The mentioned works show that the enforcement of physics constraints has the additional advantage of improving the extrapolation power of machine-learning-based models. 
An alternative stream of research avoids the fitting of a parametric material model altogether and aims to solve boundary value problems in which the material model is replaced by experimental data, resulting in so-called (material) model-free data-driven approaches (\cite{kirchdoerfer_data-driven_2016,ibanez_manifold_2018}). Here the modeling components are limited to the momentum balance equation and the assumed kinematics, and thus are reduced to a minimum. The same problem can be formulated in an inverse setting to recover stress fields from kinematic measurements, see \cite{leygue_data-based_2018,dalemat_measuring_2019,stainier_model-free_2019}.

Most of the previously discussed methods have two weak points in common. Firstly, both the training procedure for the machine learning models and the construction of a material database for the model-free methods require a large number of labeled stress-strain data pairs, which are not available from standard experiments and can only be generated through computationally expensive multiscale simulations (assuming the lower-scale geometry and material behavior are known). The second weakness is the black box nature of these methods, which makes it impossible to interpret their predictions with mathematical or physical domain knowledge.

Against the outlined background and inspired by research on discovery of ordinary or partial differential equations based on data in the physics community \citep{schmidt_distilling_2009,brunton_discovering_2016}, we recently developed a new approach for automated discovery of material models, which we denoted as EUCLID (Efficient Unsupervised Constitutive Law Identification and Discovery). Originally developed for hyperelastic materials \citep{flaschel_unsupervised_2021} (see also the independent related work by \cite{wang_inference_2021}) and subsequently extended to elastoplasticity \citep{flaschel_discovering_2022} and viscoelasticity (Marino et al., in preparation), EUCLID solves both of the aforementioned issues. In contrast to most machine-learning-based methods, EUCLID relies solely on data that can be realistically acquired with common experimental techniques, namely, full-field displacement data obtained, e.g., by Digital Image Correlation (DIC), and net reaction forces. In this sense, EUCLID is related to parameter calibration methods based on full-field displacement measurements such as the Virtual Fields Method \citep{grediac_principle_1989,pierron_virtual_2012}, see also \cite{hild_digital_2006,avril_overview_2008,roux_optimal_2020} for an overview. However, instead of calibrating the parameters of an a priori assumed material model, EUCLID automatically selects the model that leads to the best interpretation of the data starting from a very large set of candidates. Leveraging sparse regression \citep{frank_statistical_1993,tibshirani_regression_1996,brunton_discovering_2016}, it is ensured that a parsimonious and interpretable material model, i.e., one with a small number of terms and parameters, is discovered. Later works on EUCLID include a study of the problem from a Bayesian perspective for quantifying the uncertainty in the discovered models \citep{joshi_bayesian-euclid_2022} and an unsupervised training framework for artificial neural networks \citep{thakolkaran_nn-euclid_2022}. For related works that aim to identify material parameters of an a priori known constitutive model or train uninterpretable black-box machine learning models in an unsupervised fashion, the reader is referred to works by \cite{man_neural_2011,huang_learning_2020,liu_learning_2020,amores_crossing_2022,anton_identification_2022}. 

Thus far, we developed EUCLID for a few different material classes, including hyperelasticity \citep{flaschel_unsupervised_2021}, elastoplasticity \citep{flaschel_discovering_2022}) and viscoelasticity (Marino et al., in preparation). 
For each material class, we formulate a general parametric model library within the class, from which the model is automatically selected (and its unknown parameters simultaneously calibrated) based on sparse regression. This strategy is appropriate if the material class is known a priori, which may be realistic in many practical scenarios. On the other hand, in more complex situations it may well happen that the studied material is completely or largely unknown, so that the constitutive material class it belongs to is open to question. Thus, the objective of this work is to extend the scope of EUCLID to the challenging task of automatically discovering the material class along with the particular model within this class, as usual on the basis of full-field displacement and net force data. To pursue this goal, we leverage the theoretical framework of generalized standard materials.

The theory of generalized standard materials (sometimes denoted as standard dissipative materials) provides a general material modeling framework rooted in the fundamental results by \cite{onsager_reciprocal_1931-1,onsager_reciprocal_1931} on the thermodynamics of irreversible processes.
Onsager departs from a classical description of the kinetics of irreversible processes by a linear system of equations and proves the symmetry of its coefficient matrix, expressed by the so-called Onsager's reciprocal relations. This property in turn enables the definition of a quadratic dissipation potential from which the linear kinetic evolution equations can be obtained, and the establishment of a variational principle, recently also denoted as Onsager's variational principle by the soft matter community \citep{doi_onsagers_2011}.
Onsager's thermodynamic framework is extended to material modeling by \cite{ziegler_thermodynamik_1957,ziegler_attempt_1958,ziegler_systems_1972} and \cite{halphen_sur_1975} (among many others), with \cite{halphen_sur_1975} introducing the notion of generalized standard materials. Following these studies, it emerges that constitutive relations for many types of materials can be deduced from a single pair of scalar functions (thermodynamic potentials) fully characterizing the material behavior: the free energy and the dissipation potential. 
Unlike in the work by \cite{onsager_reciprocal_1931-1,onsager_reciprocal_1931}, \cite{ziegler_thermodynamik_1957} and \cite{halphen_sur_1975} consider thermodynamic potentials that are not necessarily quadratic, and hence may lead to non-linear relations between thermodynamic forces and state variable rates, but convex, which is shown to guarantee the fulfillment of thermodynamic consistency by construction. 
This generality enables the description of a large variety of different material classes with possibly non-linear constitutive equations, including elasticity, plasticity, viscosity, damage and more (see \cite{steinmann_catalogue_2021} for a recent catalogue in the one-dimensional case).

Due to the general description of the material behavior and the automatic fulfillment of physical laws, the theory of generalized standard materials constitutes an attractive framework for developing data-driven and machine-learning-based methods.
It is hence surprising that thus far little effort has been made in this direction.
One exception is the work by \cite{yu_onsagernet_2021}, who use neural networks to parameterize the system matrices, the energy and the external forces occurring in a generalized form of Onsager's reciprocal relations for modeling of the non-linear kinetics of fluids.
Another exception is the recent work by \cite{huang_variational_2022}, who propose Variational Onsager Neural Networks to parameterize the free energy potential and the dissipation potential for thermodynamic processes including phase transformations and diffusion.
The method is also applied to learn the constitutive relations of a solid material;
however, the provided benchmark test is restricted to one-dimensional viscoelasticity, with one scalar internal variable and a quadratic dissipation potential.
Thus, more complicated material behaviors with several internal variables and non-quadratic dissipation potentials (such as e.g. in plasticity) remain untackled.
A further stream of research \citep{gonzalez_thermodynamically_2019,hernandez_structure-preserving_2021,zhang_gfinns_2022} combines data-driven methods and machine learning with the GENERIC framework, proposed by \cite{grmela_dynamics_1997} and related to the theory of generalized standard materials \citep{hutter_formulation_2011,mielke_formulation_2011}.
Their objective is to identify two thermodynamic potentials and additionally (unlike with the theory of generalized standard materials) two system matrices that govern the evolution of the state variables.
The training, however, is supervised and applications to solid material modeling are thus far restricted to quadratic thermodynamic potentials, e.g., linear viscoelasticity \citep{gonzalez_learning_2019}, while once again more sophisticated material behavior remains untackled.

In the following, we provide an introduction to the theory of generalized standard materials and the resulting constitutive equations (\sectionname\ref{sec:GSM}). Subsequently, we integrate this theory in the EUCLID framework (\sectionname\ref{sec:method}), which we test on numerical benchmarks in \sectionname\ref{sec:benchmarks}. We finally draw conclusions in \sectionname\ref{sec:conclusions}.

\section{Generalized standard materials}
\label{sec:GSM}
In this section, we concisely introduce the fundamental notions on generalized standard material models, their properties, and the numerical solution of the related constitutive equations. Many more details on the models can be retrieved e.g. in the early papers by  \cite{ziegler_thermodynamik_1957,ziegler_attempt_1958} and \cite{halphen_sur_1975} or in the recent presentation by \cite{steinmann_catalogue_2021}.

\subsection{Thermodynamic potentials}
We consider a material in three-dimensional space, undergoing infinitesimal strains under isothermal conditions without phase transitions. The state of the material is assumed to be uniquely described by a set of $n+1$ \textit{state variables} $\{ \bfepsilon,\boldsymbol{\alpha}_1,\dots,\boldsymbol{\alpha}_n \}$, where $\bfepsilon$ is the (observable) infinitesimal strain tensor and $\bfalpha_i$, $i \in 1,...,n$, are internal (i.e., non-observable) state variables, typically describing the current state in relation to irreversible phenomena taking place in the material. Although $\bfalpha_i$ may be tensors of any order, they will be denoted by boldface letters in the subsequent derivations. For notational compactness we will also denote the set of internal state variables as $\bfalpha = \{ \boldsymbol{\alpha}_1,\dots,\boldsymbol{\alpha}_n \}$.

By definition, \textit{generalized standard materials} are materials for which the two following scalar functions, also called \textit{thermodynamic potentials}, exist: 
\begin{itemize}
\item The Helmholtz free energy density potential $\psi\left(\bfepsilon,\bfalpha\right)$ (in the following also simply denoted as \textit{free energy potential}), that depends on the state variables and is assumed to be strictly convex, non-negative ($\psi\left(\bfepsilon,\bfalpha\right)\geq0$), zero at the origin ($\psi\left(\bf0,\bf0\right)=0$) and smooth, i.e., continuously differentiable.
\item The dissipation rate density potential $\pi\left(\dot{\bfepsilon},\dot{\boldsymbol{\alpha}}\right)$ (or \textit{dissipation potential} in short), that depends on the time derivatives of the state variables (denoted by superposed dots), is convex, non-negative ($\pi\left(\dot{\bfepsilon},\dot{\boldsymbol{\alpha}}\right)\geq0$), zero at the origin ($\pi\left(\bf0,\bf0\right)=0$) and continuous. It may be weakly convex, i.e., locally linear, and non-smooth, i.e., not continuously differentiable.
\end{itemize}

For generalized standard materials, the specific choice of the functional form of the two thermodynamic potentials, along with the knowledge of the values of the contained parameters, completely characterizes the material behaviour. 
The properties of the free energy potential guarantee material stability and uniquely defined partial derivatives. Those of the dissipation potential will be seen to guarantee by construction the fulfillment of the second law of thermodynamics. Hence, a material model that fits within the framework of generalized standard materials is \textit{by construction stable and thermodynamically consistent}.

Note that, in general, the thermodynamic potentials could also depend on spatial gradients of the state variables, e.g., $\psi\left(\bfepsilon,\nabla \bfepsilon,\bfalpha,\nabla \bfalpha\right)$ (see \cite{nguyen_standard_2010, miehe_multi-field_2011}), or the dissipation potential could also depend on the state variables, i.e., $\pi\left(\bfepsilon,\bfalpha,\dot{\bfepsilon},\dot{\boldsymbol{\alpha}}\right)$ (see \cite{kumar_two-potential_2016}). In this initial investigation, we will consider the "classical" dependencies and will not be concerned with such cases. However, they would certainly be of interest for further generalization of the method proposed in this paper, e.g., to softening material behavior.

\subsection{State laws, complementary laws and constitutive laws}
Under the present assumptions, the so-called \textit{intrinsic dissipation} (or \textit{mechanical dissipation}) ${D}$ reads
\begin{equation}
{D}=\bfsigma\colon\dot{\bfepsilon}-\dot{\psi}\geq0,
\label{eq:D}
\end{equation}
where $\bfsigma$ is the Cauchy stress tensor and the inequality, also known as \textit{dissipation inequality}, is to be seen as a requirement posed to the material model in order to satisfy the second principle of continuum thermodynamics. Using the chain rule, the time derivative of the free energy potential can be expressed as
\begin{equation}
\label{eq:psi_dot}
\dot{\psi}=\frac{\partial\psi}{\partial\bfepsilon}\colon\dot{\bfepsilon}+\frac{\partial\psi}{\partial\boldsymbol{\alpha}_i}\colon\dot{\boldsymbol{\alpha}}_i,
\end{equation}
where we have adopted the Einstein convention for summation over repeating indices. Let us now define the \textit{reversible} or \textit{energetic stress} $\bfsigma^{en}$ and the \textit{energetic driving forces} $\boldsymbol{\mathcal{A}}^{en}_i$ as follows
\begin{equation}
\label{eq:state_laws}
\bfsigma^{en}=\frac{\partial\psi}{\partial\bfepsilon}\left(\bfepsilon,\boldsymbol{\alpha}\right),
\qquad
\boldsymbol{\mathcal{A}}^{en}_i=\frac{\partial\psi}{\partial\boldsymbol{\alpha}_i}\left(\bfepsilon,\boldsymbol{\alpha}\right).
\end{equation}
In simple terms, \textit{$\bfsigma^{en}$} represents the reversible
part of the stress, while $\boldsymbol{\mathcal{A}}^{en}_i$ is the
energetic driving force associated to the state variable $\boldsymbol{\alpha}_i$.
Equations (\eqref{eq:state_laws}) are known as \textit{state laws} and define the thermodynamic forces which are available at a given
state. By substituting \eqref{eq:psi_dot} and \eqref{eq:state_laws} in \eqref{eq:D} we obtain
\begin{equation}
\label{dissip}
{D}=(\bfsigma-\bfsigma^{en})\colon\dot{\bfepsilon}-\boldsymbol{\mathcal{A}}^{en}_i\colon\dot{\boldsymbol{\alpha}}_i=\bfsigma^{dis}\colon\dot{\bfepsilon}+\boldsymbol{\mathcal{A}}^{dis}_i\colon\dot{\boldsymbol{\alpha}}_i\geq0.
\end{equation}
where we have defined the \textit{irreversible} or \textit{dissipative
stress $\bfsigma^{dis}$ }and the \textit{dissipative driving force
}$\boldsymbol{\mathcal{A}}^{dis}_i$ as follows
\begin{equation}
\bfsigma^{dis}=\bfsigma-\bfsigma^{en},
\qquad
\boldsymbol{\mathcal{A}}^{dis}_i=-\boldsymbol{\mathcal{A}}^{en}_i\label{eq:def_dissip}.
\end{equation}

For the case of non-dissipative material behavior (${D}=0$), we obtain $\bfsigma^{dis}=\bf0$ and $\boldsymbol{\mathcal{A}}^{dis}_i=\bf0$, implying that the entire constitutive behavior is described by the state laws. For the general case ${D}\geq0$, which is considered here, the state laws are not sufficient and must be supplemented by the so-called \textit{complementary laws}
\begin{equation}
\label{eq:complementary_laws}
\bfsigma^{dis} = \frac{\partial\pi}{\partial\dot{\bfepsilon}}\left(\dot{\bfepsilon},\dot{\boldsymbol{\alpha}}\right),
\qquad
\boldsymbol{\mathcal{A}}^{dis}_i = \frac{\partial\pi}{\partial\dot{\boldsymbol{\alpha}}_i}\left(\dot{\bfepsilon},\dot{\boldsymbol{\alpha}}\right),
\end{equation}
which describe the relations between the advancement of irreversible phenomena and the corresponding thermodynamic
forces. In the case of a non-smooth dissipation potential, the partial derivatives above have to be substituted by subderivatives \citep{rockafellar_convex_2015}. In the following, for notational simplicity we denote both partial derivatives and subderivatives with the same symbol. We further introduce the set of dissipative driving forces as $\boldsymbol{\mathcal{A}}^{dis} = \{ \boldsymbol{\mathcal{A}}^{dis}_1,\dots,\boldsymbol{\mathcal{A}}^{dis}_n \}$.
Substituting \eqref{eq:complementary_laws} into \eqref{dissip} yields
\begin{equation}
{D}=\frac{\partial\pi}{\partial\dot{\bfepsilon}}\colon\dot{\bfepsilon} + \frac{\partial\pi}{\partial\dot{\boldsymbol{\alpha}}_i}\colon\dot{\boldsymbol{\alpha}}_i.
\end{equation}
It is thus evident that, whatever the specific choice of $\pi$ with the postulated properties, it is $D \geq 0$, i.e., the dissipation inequality is satisfied by construction. 

The combination of state laws and complementary laws is sufficient to completely describe the material behavior, which is governed by the constitutive laws
\begin{equation}
\label{eq:constitutive_laws}
\bfsigma = \frac{\partial\psi}{\partial\bfepsilon}\left(\bfepsilon,\boldsymbol{\alpha}\right) + \frac{\partial\pi}{\partial\dot{\bfepsilon}}\left(\dot{\bfepsilon},\dot{\boldsymbol{\alpha}}\right),
\qquad
\bm{0} = \frac{\partial \psi}{\partial \bfalpha_i}\left(\bfepsilon,\boldsymbol{\alpha}\right) + \frac{\partial \pi}{\partial \dot{\bfalpha}_i}\left(\dot{\bfepsilon},\dot{\boldsymbol{\alpha}}\right).
\end{equation}

\subsection{Dual dissipation potential and dual complementary laws}
An equivalent alternative to the complementary laws is obtained by applying the Legendre-Fenchel transformation\footnote{As the dissipation potential might not be strictly convex and smooth, the conventional Legendre transformation is not applicable, see \cite{touchette_legendre-fenchel_2007} for details.} to the dissipation potential to obtain the \textit{dual dissipation potential}
\begin{equation}
\pi^{*}\left(\bfsigma^{dis},\boldsymbol{\mathcal{A}}^{dis}\right)=\max_{\dot{\bfepsilon},\mathbf{\dot{\boldsymbol{\alpha}}}}\left\{ \bfsigma^{dis}\colon\dot{\bfepsilon}+\boldsymbol{\mathcal{A}}^{dis}_i\colon\dot{\boldsymbol{\alpha}}_i-\pi\left(\dot{\bfepsilon},\dot{\boldsymbol{\alpha}}\right)\right\},
\end{equation}
from which we obtain
\begin{equation}
\label{eq:dual_complementary_laws}
\dot{\bfepsilon} = \frac{\partial\pi^*}{\partial\bfsigma^{dis}}\left(\bfsigma^{dis},\boldsymbol{\mathcal{A}}^{dis}\right),
\qquad
\dot{\boldsymbol{\alpha}}_i = \frac{\partial\pi^*}{\partial\boldsymbol{\mathcal{A}}^{dis}_i}\left(\bfsigma^{dis},\boldsymbol{\mathcal{A}}^{dis}\right),
\end{equation}
known as \textit{dual complementary laws}. These laws describe
the advancement of irreversible phenomena as a function of the corresponding
thermodynamic forces, i.e., they are basically evolution equations
for the strain and for the internal variables, which is why they are
often more useful than \eqref{eq:complementary_laws}. Once again, should
$\pi^*$ be non-smooth, the partial derivatives have to be intended as sub-derivatives. Note that, due to a property of the Legendre-Fenchel transform, convexity of $\pi$ is equivalent to that of $\pi^*$.

\subsection{Strain-rate-independent dissipation potential}
For solid materials, it is often assumed that the dissipation potential does not depend on the strain rate, i.e., $\pi=\pi\left(\dot{\boldsymbol{\alpha}}\right)$ (see, e.g., \cite{lemaitre_mechanics_1994,michel_model-reduction_2016,bluhdorn_automat_2022}).  As a consequence, it is $\pi^*=\pi^*\left(\boldsymbol{\mathcal{A}}^{dis}\right)$ and the constitutive laws can be simplified to
\begin{equation}
\label{eq:constitutive_laws_simple}
\bfsigma=\frac{\partial\psi}{\partial\bfepsilon}\left(\bfepsilon,\boldsymbol{\alpha}\right),
\qquad
\dot{\boldsymbol{\alpha}}_i = \left. \frac{\partial\pi^*}{\partial\boldsymbol{\mathcal{A}}^{dis}_i}\left(\boldsymbol{\mathcal{A}}^{dis}\right) \right\rvert_{\boldsymbol{\mathcal{A}}^{dis}_i = -\frac{\partial\psi}{\partial{\boldsymbol{\alpha}}_i}}.
\end{equation}
In this work, we consider a dissipation potential which is independent of the strain rate. However, only minor modifications would be needed to extend the framework developed in the following to the case of a strain-rate-dependent dissipation potential. Note that the choice of a strain-rate-independent dissipation potential does not imply a strain-rate-independent material model, as rate dependency is kept in the evolution of the internal variables.

\subsection{Numerical solution of the constitutive equations}
Given an initial condition for the state variables, \eqnames\eqref{eq:constitutive_laws_simple} can be solved either in strain or in stress control for the history of the state variables and of the dependent variables. Here, we will consider strain control, i.e., the strain history is given and the objective is to compute the stress and the internal variables at each time. The conventional strategy to solve such problems is to compute the partial derivatives (or subderivatives) of the thermodynamic potentials analytically and to solve at each time the resulting equations, which are non-linear in general, using e.g. the Newton-Raphson method. If the dissipation potential is non-smooth, algorithms like the return mapping in elastoplasticity \citep{simo_computational_1998,neto_computational_2008} are commonly applied.

Alternative strategies for solving equations like \eqref{eq:constitutive_laws_simple} employ automatic differentiation, see \cite{korelc_multi-language_2002,rothe_automatic_2015,bluhdorn_automat_2022}. In this way, derivatives do not need to be evaluated by hand and the process of implementing the constitutive equation solver for new models is greatly simplified, which can be particularly beneficial during the development stage of new material models. However, such automatic solution strategies also have drawbacks. For non-smooth thermodynamic potentials, automatic differentiation cannot be applied, thus the potentials must be approximated by a continuously differentiable regularization (examples are provided by \cite{steinmann_catalogue_2021}), whose choice is user-dependent and does not follow rigorous rules. The regularization also introduces additional parameters and non-linearity to the problem. Further, while conventional methods often take advantage of problem-specific simplifications that increase the computational efficiency (see, e.g., the scalar internal variable substitution during the viscoplastic corrector step in \ref{sec:stress_update}), automatic methods are blind to such possibilities, thus they tend to be computationally more expensive. For these reasons, in this work we rely on the conventional methods for solving the constitutive equations and for computing the consistent tangent (see \ref{sec:stress_update} and \ref{sec:consisten_tangent} for details).

\section{Automated discovery of generalized standard material models}
\label{sec:method}
In the following, we describe our novel approach to automatically discover generalized standard (hence, stable and thermodynamically consistent) material models based on full-field displacement and global force data. The backbone of the approach, which we denote as EUCLID in line with our previous work \citep{flaschel_unsupervised_2021,flaschel_discovering_2022,joshi_bayesian-euclid_2022,thakolkaran_nn-euclid_2022}, can be outlined as follows: we start from a very wide material model space, which we construct by exploiting the flexibility and a priori thermodynamic consistency of the generalized standard material model framework; we constrain this model space by enforcing the satisfaction of momentum balance (in weak sense) on the source data; we exploit sparse regression to ensure parsimonity of the discovered model. Through these three key ingredients, we end up simultaneously performing model selection and identification of the unknown material parameters. Unlike in our previous papers, our model selection here is not restricted to taking place within a predefined category of material behavior (such as elasticity or plasticity), but the category itself is one of the outcomes of the approach. 

\subsection{Material model library}
\label{sec:model_library}
\begin{figure}[htb]
\begin{center}
\includegraphics[width=0.5\textwidth]{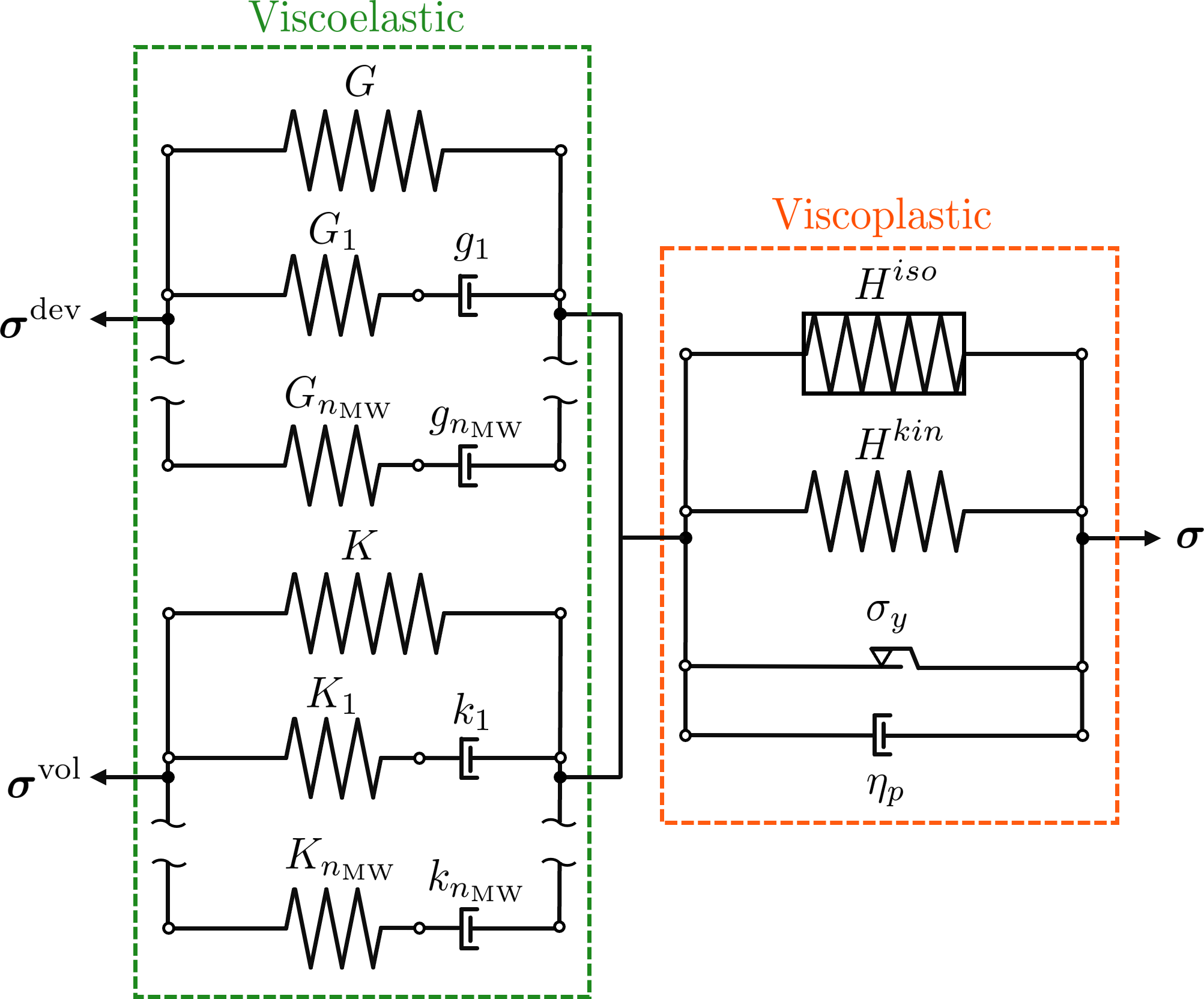}
\caption{Rheological representation of the material model library. For the interpretation of the isotropic hardening element see \cite{brocker_thermoviscoplasticity_2012}.}
\label{fig:rheo}
\end{center}
\end{figure}

We start by constructing a parametric library of three-dimensional and isotropic generalized standard material models, out of which the EUCLID algorithm will later select a model that best fits the data while at the same time being mathematically concise. In particular, we choose a combination of generalized Maxwell viscoelastic models and viscoplastic models including isotropic and kinematic hardening mechanisms, see \figurename\ref{fig:rheo} for a rheological illustration. Thus, the model library includes a large portion of the models included in the catalogue by \cite{steinmann_catalogue_2021}\footnote{In the catalogue by \cite{steinmann_catalogue_2021}, the models are presented in one-dimensional form.}. For similar viscoelastic-viscoplastic modeling approaches, the reader is referred to \cite{giunta_onedimensional_2006,miled_coupled_2011,miled_coupled_2011-1,sun_serial_2013}.

\subsubsection{Thermodynamic potentials}
\label{sec:model_library_potentials}
The adopted model library for the free energy potential $\psi$ reads
\begin{equation}
\begin{aligned}
\psi\left(\bfepsilon,\boldsymbol{\alpha}\right) 
&= \frac{1}{2} \left( 2G \dev(\bfepsilon - \boldsymbol{\alpha}_{\RomanNumeralCaps{1}}) \colon \dev(\bfepsilon - \boldsymbol{\alpha}_I) + K \tr^2(\bfepsilon - \boldsymbol{\alpha}_{\RomanNumeralCaps{1}}) \right) + \frac{1}{2} H^{iso} \alpha_{\RomanNumeralCaps{2}}^2 + \frac{1}{2} H^{kin} \boldsymbol{\alpha}_{\RomanNumeralCaps{3}} \colon \boldsymbol{\alpha}_{\RomanNumeralCaps{3}} \\
&+ \frac{1}{2} \sum_{j = 1}^{n_{\text{MW}}} \left( 2G_j \dev(\bfepsilon - \boldsymbol{\alpha}_j - \bfalpha_{\RomanNumeralCaps{1}}) \colon \dev(\bfepsilon  - \boldsymbol{\alpha}_j - \bfalpha_{\RomanNumeralCaps{1}}) + K_j \tr^2(\bfepsilon - \boldsymbol{\alpha}_j - \bfalpha_{\RomanNumeralCaps{1}}) \right),
\end{aligned}
\end{equation}
where $\bfalpha=\{ \bfalpha_1,\dots,\bfalpha_{n_{\text{MW}}},\bfalpha_{\RomanNumeralCaps{1}},\alpha_{\RomanNumeralCaps{2}},\bfalpha_{\RomanNumeralCaps{3}} \}$ are the $n_{\text{MW}}+3$ internal variables. Here, $n_{\text{MW}}$ denotes the number of viscoelastic Maxwell elements considered in the model library; $\bfalpha_1,\dots,\bfalpha_{n_{\text{MW}}}$ have the meaning of viscous strains in the Maxwell elements; $\bfalpha_I$ is the plastic strain tensor; $\bfalpha_{II}$ and $\bfalpha_{III}$ are the two hardening variables corresponding to isotropic and kinematic hardening; $G$ and $K$ are the long-term shear and bulk moduli; $G_i$ and $K_i$ are the shear and bulk moduli of the $i^{th}$ viscoelastic Maxwell element; and $H^{iso}$ and $H^{kin}$ are material parameters related to isotropic and kinematic hardening. At this point, the following considerations are in order:
\begin{itemize}
\item The actual number of internal variables needed for the accurate description of the material behavior is not known a priori. Our proposed strategy here consists in introducing in the library a number of such variables deemed to be larger than the actual number of internal variables needed to describe the material state. As will become evident later, unnecessary internal variables will become automatically inactive through the sparsity regularization.
\item Compared to our previous investigations focusing on a single material class, here we are limiting ourselves to a reduced library per class. E.g., unlike in \cite{flaschel_unsupervised_2021}, we are only considering linear elasticity within the kinematically linear framework, and in contrast to \cite{flaschel_discovering_2022}, we fix the shape of the initial yield surface and introduce a less versatile hardening description. The reason is that here the focus lies on the ability of EUCLID to discriminate among different material classes, rather than on its versatility within a given material class (which was extensively proved in our previous papers). Obviously, further extensions are possible and may well be addressed in future developments.
\end{itemize}

The model library for the dual dissipation potential $\pi^{*}$ reads
\begin{equation}
\pi^{*}\left(\boldsymbol{\mathcal{A}}^{dis}\right) = \pi^{*}_{\text{VE}}\left(\boldsymbol{\mathcal{A}}^{dis}_1,\dots,\boldsymbol{\mathcal{A}}^{dis}_{n_{\text{MW}}}\right) + \pi^{*}_{\text{VP}}\left(\boldsymbol{\mathcal{A}}^{dis}_{\RomanNumeralCaps{1}},\mathcal{A}^{dis}_{\RomanNumeralCaps{2}},\boldsymbol{\mathcal{A}}^{dis}_{\RomanNumeralCaps{3}}\right),
\end{equation}
with a viscoelastic contribution
\begin{equation}
\pi^{*}_{\text{VE}}\left(\boldsymbol{\mathcal{A}}^{dis}_1,\dots,\boldsymbol{\mathcal{A}}^{dis}_{n_{\text{MW}}}\right)=
\frac{1}{2} \sum_{j = 1}^{n_{\text{MW}}} \left( \frac{1}{2G_j g_j} \dev(\boldsymbol{\mathcal{A}}^{dis}_j) \colon \dev(\boldsymbol{\mathcal{A}}^{dis}_j) + \frac{1}{9K_j k_j} \tr^2(\boldsymbol{\mathcal{A}}^{dis}_j) \right),
\end{equation}
and a viscoplastic contribution
\begin{equation}
\pi^{*}_{\text{VP}}\left(\boldsymbol{\mathcal{A}}^{dis}_{\RomanNumeralCaps{1}},\mathcal{A}^{dis}_{\RomanNumeralCaps{2}},\boldsymbol{\mathcal{A}}^{dis}_{\RomanNumeralCaps{3}}\right)=
\begin{cases}
0
& \text{if} ~ \sqrt{\frac{3}{2}}\left\|\dev\left(\boldsymbol{\mathcal{A}}^{dis}_{\RomanNumeralCaps{1}} + \boldsymbol{\mathcal{A}}^{dis}_{\RomanNumeralCaps{3}}\right)\right\|\leq\sigma_{y}-\mathcal{A}^{dis}_{\RomanNumeralCaps{2}}\\
\frac{1}{2}\frac{1}{\eta_p} \left( \sqrt{\frac{3}{2}}\left\|\dev\left(\boldsymbol{\mathcal{A}}^{dis}_{\RomanNumeralCaps{1}} + \boldsymbol{\mathcal{A}}^{dis}_{\RomanNumeralCaps{3}}\right)\right\| - \sigma_{y} + \mathcal{A}^{dis}_{\RomanNumeralCaps{2}} \right)^2
& \text{otherwise},
\end{cases}
\end{equation}
where $\boldsymbol{\mathcal{A}}^{dis}=\{ \boldsymbol{\mathcal{A}}^{dis}_1,\dots,\boldsymbol{\mathcal{A}}^{dis}_{n_{\text{MW}}},\boldsymbol{\mathcal{A}}^{dis}_{\RomanNumeralCaps{1}},\mathcal{A}^{dis}_{\RomanNumeralCaps{2}},\boldsymbol{\mathcal{A}}^{dis}_{\RomanNumeralCaps{3}} \}$ are the dissipative driving forces associated to the internal variables with the same subscripts, $g_i$ and $k_i$ are the shear and bulk relaxation times of the $i^{th}$ viscoelastic Maxwell element, $\sigma_0$ is the yield stress, and $\eta_p$ is the viscosity associated to the viscoplastic behavior.

\subsubsection{Constitutive equations}
\label{sec:model_library_constitutive_laws}

The constitutive equations for the chosen model library are obtained by substituting the thermodynamic potentials of \sectionname\ref{sec:model_library_potentials} in \eqref{eq:constitutive_laws_simple}. The stress is derived as
\begin{equation}
\label{eq:stress}
\begin{aligned}
\bfsigma=\frac{\partial\psi}{\partial\bfepsilon}
&= 2G \dev(\bfepsilon - \boldsymbol{\alpha}_{\RomanNumeralCaps{1}}) + K \tr(\bfepsilon - \boldsymbol{\alpha}_{\RomanNumeralCaps{1}}) \bfI\\
&+ \sum_{j = 1}^{n_{\text{MW}}} \left( 2G_j \dev(\bfepsilon - \boldsymbol{\alpha}_j - \boldsymbol{\alpha}_{\RomanNumeralCaps{1}}) + K_j \tr(\bfepsilon - \boldsymbol{\alpha}_j - \boldsymbol{\alpha}_{\RomanNumeralCaps{1}}) \bfI \right). \\
\end{aligned}
\end{equation}

Differentiation of the dual dissipation potential with respect to the viscoelastic driving forces leads to the viscoelastic evolution laws
\begin{equation}
\dot{\boldsymbol{\alpha}}_j=\frac{\partial\pi^*}{\partial\boldsymbol{\mathcal{A}}^{dis}_j}=
\frac{1}{2G_j g_j} \dev(\boldsymbol{\mathcal{A}}^{dis}_j) + \frac{1}{9K_j k_j} \tr(\boldsymbol{\mathcal{A}}^{dis}_j) \bfI.
\end{equation}
These laws need to be evaluated at
\begin{equation}
\boldsymbol{\mathcal{A}}^{dis}_j
= -\frac{\partial\psi}{\partial\boldsymbol{\alpha}_j}
= \left( 2G_j \dev(\bfepsilon - \boldsymbol{\alpha}_j - \boldsymbol{\alpha}_{\RomanNumeralCaps{1}}) + K_j \tr(\bfepsilon - \boldsymbol{\alpha}_j - \boldsymbol{\alpha}_{\RomanNumeralCaps{1}}) \bfI \right),
\end{equation}
which leads to
\begin{equation}
\label{eq:evolution_VE}
\dot{\boldsymbol{\alpha}}_j
= \frac{1}{g_j} \dev(\bfepsilon - \boldsymbol{\alpha}_j - \boldsymbol{\alpha}_{\RomanNumeralCaps{1}}) + \frac{1}{3 k_j} \tr(\bfepsilon - \boldsymbol{\alpha}_j - \boldsymbol{\alpha}_{\RomanNumeralCaps{1}}).
\end{equation}

Differentiation of the dual dissipation potential with respect to the viscoplastic driving forces leads to the viscoplastic evolution laws
\begin{equation}
\begin{aligned}
\dot{\boldsymbol{\alpha}}_{\RomanNumeralCaps{1}} = \dot{\boldsymbol{\alpha}}_{\RomanNumeralCaps{3}}
=\frac{\partial\pi^*}{\partial\boldsymbol{\mathcal{A}}^{dis}_{\RomanNumeralCaps{1}}}
&=
\begin{cases}
0 & \text{if} ~ f \leq 0\\
\frac{1}{\eta_p} f \sqrt{\frac{3}{2}} \frac{\dev\left(\boldsymbol{\mathcal{A}}^{dis}_{\RomanNumeralCaps{1}} + \boldsymbol{\mathcal{A}}^{dis}_{\RomanNumeralCaps{3}}\right)}{\left\|\dev\left(\boldsymbol{\mathcal{A}}^{dis}_{\RomanNumeralCaps{1}} + \boldsymbol{\mathcal{A}}^{dis}_{\RomanNumeralCaps{3}}\right)\right\|} & \text{otherwise},
\end{cases}\\
\dot{{\alpha}}_{\RomanNumeralCaps{2}}
=\frac{\partial\pi^*}{\partial{\mathcal{A}}^{dis}_{\RomanNumeralCaps{2}}}
&=
\begin{cases}
0 & \text{if} ~ f \leq 0\\
\frac{1}{\eta_p} f & \text{otherwise},
\end{cases}
\end{aligned}
\end{equation}
where we defined
\begin{equation}
f = \sqrt{\frac{3}{2}}\left\|\dev\left(\boldsymbol{\mathcal{A}}^{dis}_{\RomanNumeralCaps{1}} + \boldsymbol{\mathcal{A}}^{dis}_{\RomanNumeralCaps{3}}\right)\right\| - \sigma_{y} + \mathcal{A}^{dis}_{\RomanNumeralCaps{2}}.
\end{equation}
We notice that $\dot{\boldsymbol{\alpha}}_{\RomanNumeralCaps{1}} = \dot{\boldsymbol{\alpha}}_{\RomanNumeralCaps{3}}$. These evolution laws need to be evaluated at $\boldsymbol{\mathcal{A}}^{dis}_{\RomanNumeralCaps{1}} = -\frac{\partial\psi}{\partial\boldsymbol{\alpha}_{\RomanNumeralCaps{1}}} = \bfsigma$, ${\mathcal{A}}^{dis}_{\RomanNumeralCaps{2}} = -\frac{\partial\psi}{\partial{\alpha}_{\RomanNumeralCaps{2}}} = -H^{iso}\alpha_{\RomanNumeralCaps{2}}$ and $\boldsymbol{\mathcal{A}}^{dis}_{\RomanNumeralCaps{3}} = -\frac{\partial\psi}{\partial\boldsymbol{\alpha}_{\RomanNumeralCaps{3}}} = -H^{kin}\boldsymbol{\alpha}_{\RomanNumeralCaps{3}}$, leading to
\begin{equation}
\label{eq:evolution_VP}
\begin{aligned}
\dot{\boldsymbol{\alpha}}_{\RomanNumeralCaps{1}} = \dot{\boldsymbol{\alpha}}_{\RomanNumeralCaps{3}}
&=
\begin{cases}
0 & \text{if} ~ f \leq 0\\
\frac{1}{\eta_p} f \sqrt{\frac{3}{2}} \frac{\dev\left(\bfsigma -H^{kin}\boldsymbol{\alpha}_{\RomanNumeralCaps{3}}\right)}{\left\|\dev\left(\bfsigma -H^{kin}\boldsymbol{\alpha}_{\RomanNumeralCaps{3}}\right)\right\|} & \text{otherwise},
\end{cases}\\
\dot{{\alpha}}_{\RomanNumeralCaps{2}}
&=
\begin{cases}
0 & \text{if} ~ f \leq 0\\
\frac{1}{\eta_p} f & \text{otherwise},
\end{cases}
\end{aligned}
\end{equation}
with
\begin{equation}
f = \sqrt{\frac{3}{2}}\left\|\dev\left(\bfsigma -H^{kin}\boldsymbol{\alpha}_{\RomanNumeralCaps{3}}\right)\right\| - \sigma_{y} -H^{iso}\alpha_{\RomanNumeralCaps{2}}.
\end{equation}

For solving the incremental constitutive equations at a material point, we choose an implicit Euler discretization in time. Given the state variables at the previous time step, the strain at the current time step and the time step size, a viscoelastic predictor / viscoplastic corrector algorithm (see \ref{sec:stress_update}) is applied to compute the stress and the internal variables at the current time step. 
The consistent tangent modulus is given in \ref{sec:consisten_tangent}.

\subsubsection{Material parameters}
\label{sec:model_library_parameters}
All tunable material parameters in the material model library are collected in the material parameter vector
\be
\label{eq:theta_vector}
\bftheta = 
\left[
G,K,G_1,\dots,G_{n_{\text{MW}}},\frac{1}{g_1},\dots,\frac{1}{g_{n_{\text{MW}}}},K_1,\dots,K_{n_{\text{MW}}},\frac{1}{k_1},\dots,\frac{1}{k_{n_{\text{MW}}}},\frac{1}{\sigma_0},\eta_p,H^{iso},H^{kin}
\right]^T.
\ee
To ensure the convexity of the thermodynamic potentials, all parameters in $\bftheta$ are assumed to be non-negative.

Note that some parameters appear in the material parameter vector through their reciprocals. This facilitates the simplification of the model in special cases for which these parameters tend to infinity (see also \sectionname\ref{sec:sparsity}). For example, if the yield stress tends to infinity, the model can be considered as purely viscoelastic, and if a relaxation time tends to infinity, the viscous effect of the corresponding Maxwell element can be ignored. In practice, it is more convenient to work with vanishing quantities than with diverging ones. For the numerical treatment, the parameters in reciprocal form are assumed to be strictly positive.

\subsection{Available data}

By conducting a mechanical test on a specimen made of the material under consideration, two-dimensional full-field displacements and global reaction force data can be acquired to serve as input for EUCLID. The point-wise displacement data are given in the form $\{\bfu^{a,t}:a=1,\dots,n_n; t=1,\dots,n_t\}$, where $n_n$ is the number of points in space and $n_t$ is the number of time steps at which displacements are known. The reaction force data are given in the form $\{\hat R^{\beta,t}:\beta=1,\dots,n_\beta; t=1,\dots,n_t\}$, where $n_\beta$ is the number of measured reaction force components.

To interpolate the given displacement data, we create a finite element mesh with shape functions $\{N^a(\bfX):a=1,\dots,n_n\}$, such that each node of the mesh corresponds to one of the points where the displacement is known. The displacement field at each time is then given as
\begin{equation}
\bfu^t(\bfX) = \sum_{a=1}^{n_n} N^a(\bfX)\bfu^{a,t}, \quad \forall \ t=1,\dots,n_t,
\end{equation}
which is differentiated in space to obtain the infinitesimal strain field at each time step
\be
\bfeps^\idxLOAD(\bfX)
= \nabla^{\text{sym}}\bfu^\idxLOAD(\bfX)
= \sum_{a=1}^{n_n} \frac{1}{2} \left[\bfu^{a,\idxLOAD} \otimes \nabla N^a(\bfX) + \nabla N^a(\bfX) \otimes \bfu^{a,\idxLOAD}\right], \quad \forall \ t=1,\dots,n_t.
\ee

\subsection{Cost function based on the weak linear momentum balance}
\label{sec:cost_function}

Given full-field kinematic data and net reaction forces, EUCLID employs the balance of linear momentum as a physical constraint to determine the optimal values of the unknown parameters in the model library. In our previous investigations, we observed this constraint to be effective in contrasting the strong ill-posedness induced by the lack of labeled stress-strain data pairs. In the following, we introduce a cost function based on the sum of squared unbalanced forces, thus establishing the weak linear momentum balance in the interior and at the boundary of the test specimen (see \cite{flaschel_unsupervised_2021,flaschel_discovering_2022} for more information). Later, this cost function will be minimized along with a sparsity promoting penalty term to select a parsimonious model while determining its unknown parameters.

The weak formulation of the linear momentum balance under the assumption of vanishing inertia and body forces at a generic time step $t$ reads
\be\label{eq:weak_form}
\int_\Omega \bfsigma^\idxLOAD(\Delta t,\bfeps^\idxLOAD,\bfalpha^{\idxLOAD-1},\bftheta)\colon\nabla \bfv \dd A - \int_{\partial\Omega}\hat \bft^\idxLOAD \cdot \bfv \dd S = 0, \quad \forall \  \text{admissible} \ \bfv, \quad \forall \ t=1,\dots,n_t,
\ee
where $\Omega$ and $\partial\Omega$ denote the specimen domain and boundary, respectively, and $\hat \bft$ is the surface traction on $\partial\Omega$. The stress at the considered time step $\bfsigma^\idxLOAD$ depends on the time step size $\Delta t$, the current strain $\bfeps^\idxLOAD$, the internal variables at the previous time step $\bfalpha^{\idxLOAD-1}$ and the material parameters $\bftheta$ through the stress update algorithm (see \ref{sec:stress_update}). The weak form must be satisfied for any admissible (i.e., sufficiently smooth) test functions $\bfv$; we discretize them with the same shape functions adopted for the displacement data interpolation 
\be
\bfv(\bfX) = \sum_{a=1}^{n_n}N^a(\bfX) \,\bfv^a,
\ee
and we assume them to be constant in time. This leads to the discretized weak linear momentum balance
\be\label{eq:weak_form_discrete}
\sum_{a=1}^{n_n}\bfv^a\cdot\left[ \underbrace{\int_\Omega \bfsigma^\idxLOAD(\Delta t,\bfeps^\idxLOAD,\bfalpha^{\idxLOAD-1},\bftheta)\nabla N^a(\bfX) \dd A}_{\bfF^{a,t}(\bftheta)} - \int_{\partial\Omega} {\hat\bft}^t N^a(\bfX) \dd S\right] = 0, \quad \forall \  \text{admissible} \ \bfv, \quad \forall \ t=1,\dots,n_t,
\ee
where $\bfF^{a,t}$ denotes the internal force at node $a$ of the finite element mesh for time step $t$. Since the internal forces depend on the stress at the current time step, which in turn depends on the internal variables at the previous time step, they can only be computed (for a given set of material parameters $\bftheta$) by iterating through all load steps and solving the constitutive equations at each step. To this end, the internal variables at the initial time step are assumed to be zero.

We denote the set of all nodal degrees of freedom in the finite element mesh as $\calD=\{(a,i) : a=1,\dots,n_n;i=1,2\}$. We further partition this set into two subsets: $\calD^{\text{free}}$, containing the free degrees of freedom, i.e., those that do not correspond to any of the constrained boundaries, and $\calD^{\text{disp}}$, containing the displacement-constrained degrees of freedom, such that $\calD^{\text{free}}\cup\calD^{\text{disp}}=\calD$ and $\calD^{\text{free}}\cap\calD^{\text{disp}}=\emptyset$. Further, we denote the subset of $\calD^{\text{disp}}$ corresponding to the reaction force $\hat R^{\beta,t}$ as $\calD^{\text{disp},\beta}$.

At the free degrees of freedom, the traction force $\hat \bft$ vanishes and we obtain from \eqref{eq:weak_form_discrete}
\be\label{eq:weak_free}
F^{a,t}_i(\bftheta) = 0, \quad \forall \ (a,i)\in\calD^\text{free}, \quad \forall \ t=1,\dots,n_t,
\ee
whereas at the displacement-constrained boundaries the sum of the nodal internal forces must equal the global external reaction force
\be\label{eq:weak_disp}
\sum_{(a,i)\in\calD^{\text{disp},\beta}}F^{a,t}_i(\bftheta) = \sum_{(a,i)\in\calD^{\text{disp},\beta}} \int_{\partial\Omega} {\hat t}^t_i N^a(\bfX) \dd S = \hat R^{\beta,t}, \quad \forall \ \beta=1,\dots,n_{\beta}, \quad \forall \ t=1,\dots,n_t.
\ee
Global measures of the violation of linear momentum balance in the interior and at the boundary can thus be obtained by the following sums of squared residuals
\be\label{eq:Cfree}
C^\text{free}(\bftheta) = \sum_{t=1}^{n_t}\sum_{(a,i) \in \calD^\text{free}} \left|F^{a,t}_i(\bftheta)\right|^2,
\ee
and
\be\label{eq:Cdisp}
C^\text{disp}(\bftheta) = \sum_{t=1}^{n_t} \sum_{\beta=1}^{n_\beta} \left|\hat R^{\beta,t} - \sum_{(a,i)\in \calD^{\text{disp},\beta} }  F^{a,t}_i(\bftheta) \right|^2,
\ee
which are finally combined in the 
cost function
\be\label{eq:cost_function}
C(\bftheta) =
C^{\text{free}}(\bftheta)
+
\lambda_r
C^{\text{disp}}(\bftheta).
\ee
Here $\lambda_r$ is a weighting parameter that scales the interior and boundary contributions to the cost function. As there are less reaction force measurements (see \eqref{eq:weak_disp}) than free degrees of freedom (see \eqref{eq:weak_free}), the weighting parameter should be chosen sufficiently larger than one ($\lambda_r>>1$). Following previous works \citep{flaschel_unsupervised_2021, flaschel_discovering_2022}, we choose $\lambda_r = 100$ and keep it constant throughout this work. Based on our experience, the choice of $\lambda_r$ is not crucial for the success of the method. Finally, note that in the quantification of the cost we are summing upon all time steps.

\subsection{Sparsity promoting regularization}
\label{sec:sparsity}

Material models with a large number of material parameters are known to be less interpretable and
poor at extrapolation to unseen strains \citep{hartmann_numerical_2001}. For this reason, our objective with EUCLID is to obtain a parsimonious model, i.e., a model with as few features as needed to accurately interpret the available data. Minimizing the cost function in \eqref{eq:cost_function} leads in general to a dense parameter vector $\bftheta$, i.e., a vector with many non-zero entries. This, in turn, implies that the resulting material model contains nearly all terms of the thermodynamic potentials contained in the initially chosen library (see \sectionname\ref{sec:model_library}). Thus, nearly all internal variables are active and the calibrated model is as complex as the entire model space adopted at the outset,  which goes against the objective of discovering an interpretable model containing only a few terms. In order to obtain a sparse parameter vector with many zero entries and hence simple expressions for the thermodynamic potentials, we apply sparse regression, i.e., we minimize \eqref{eq:cost_function} supplemented with a sparsity promoting regularization term
\be
\label{eq:objective_regularized}
\bftheta^{\text{opt}} = 
\argminWithArgs_{\bftheta\geq\bftheta^{\text{min}}}
\left(
C(\bftheta)
+ \lambda_p\|\bftheta \|_p^p
\right),
\qquad \text{where} \qquad 
\|\bftheta\|_p = \left(\sum_{i \geq 3}|\theta_i|^p\right)^{1/p}.
\ee
The sparsity promoting regularization term was originally proposed by \cite{frank_statistical_1993} and \cite{tibshirani_regression_1996} and later applied to problems in dynamics by \cite{brunton_discovering_2016} and to problems in continuum mechanics by \cite{flaschel_unsupervised_2021,flaschel_discovering_2022}.
This term takes high values for dense solution vectors and small values for sparse solution vectors, whereby the weighting parameter $\lambda_p$ influences its impact on the minimization problem.
We will show later how the value of $\lambda_p$ can be automatically selected to achieve a user-defined compromise between model accuracy and sparsity (see \sectionname\ref{sec:hyperparameters}).
The parameter $p$ influences the curvature of the regularization term.
For $p \rightarrow 0$, this term converges to the $L_0$-(pseudo)-norm of $\bftheta$, which counts the number of non-zero entries in the vector.
However, this choice turns the optimization problem into a combinatorial subset selection problem which becomes computationally intractable for a large number of features.
For this reason, $L_0$ regularization is often relaxed to the convex $L_1$ regularization ($p = 1$), also known as LASSO (least absolute shrinkage and selection operator) \citep{tibshirani_regression_1996}.
As the $L_0$ regularization, the LASSO promotes sparse solution vectors with many zero entries, while shrinking the remaining non-zero parameters.
Within $[0, 1]$, a smaller value of $p$ reduces the shrinkage of the non-zero parameters at the cost of computational complexity, as the degree of non-convexity in the penalty term increases with decreasing $p$.
In this work, we thus choose $p=1$ to limit the computational complexity, and keep it constant throughout the numerical examples.
Based on our experience, the choice of $0 < p \leq 1$ is not crucial for the success of the method \citep{flaschel_unsupervised_2021,flaschel_discovering_2022}.

As different material parameters have different units, the summation in the regularization term is unphysical and only fulfills a numerical purpose. While a non-dimensionalization strategy would be recommendable (and is planned for future studies), here we adopt the simplified approach to compute the regularization term based on the numerical values of the parameters for fixed units (chosen as $\text{s}$ for time, $\text{mm}$ for length and $\text{kN}$ for force) and treat the result as dimensionless, such that the unit of the weighting factor $\lambda_p$ is assumed to be $\text{kN}^2$. With this specific choice of units, the numerical values for the material parameters studied in this work appear within a reasonable range of orders of magnitude, such that no scaling issues were observed during numerical testing.
Note that we exclude $\theta_1$ and $\theta_2$ from the sparsity promoting regularization as we assume that the elasticity constants $G$, $K$ are always non-zero for a solid material.
The parameter constraints $\bftheta>\bftheta^{\text{min}}$ in \eqref{eq:objective_regularized} enforce that all parameters are larger than zero, and all parameters that appear in the minimization problem with their reciprocals (i.e., $1/g_i$, $1/k_i$, $1/\sigma_0$) are larger than a lower bound, here chosen as $\theta^{\text{lb}}=10^{-6}$ (see also \sectionname\ref{sec:model_library_parameters}).


\subsection{Optimization strategy and hyperparameter selection}
\label{sec:hyperparameters}
In the following, we describe the strategies adopted for the solution of the optimization problem in \eqref{eq:objective_regularized} and for the selection of the hyperparameter $\lambda_p$ (for a summary of all other (hyper-)parameters see \ref{sec:settings}). For the optimization, we rely on the trust region reflective Newton algorithm implemented in the Matlab built-in function \textit{lsqnonlin}, which has proven to be a suitable choice in terms of convergence and efficiency in previous work \citep{flaschel_discovering_2022}.

To obtain an initial estimate of the solution, the problem is first solved without regularization, i.e., $\lambda_p=0$, and for a reduced set of parameters, i.e., the linear elastic constants $G$ and $K$, while constraining the other parameters to be equal to the corresponding values in $\bftheta^{\text{min}}$. This preconditioning step is computationally efficient due to the small number of trainable parameters, and at the same time it provides a good estimate about the order of magnitude of the elastic parameters. Subsequently, the optimization problem is solved (still with $\lambda_p=0$) for the entire set of unknown parameters. In general, the function to be minimized in \eqref{eq:objective_regularized} is non-convex and the minimization problem admits multiple local minima. In order to increase the chance of finding (or approaching as closely as possible) the global optimum, we run the optimization in parallel for multiple randomly generated initial guesses ($n_g=24$ in this work) and, from all generated solutions, we select the one corresponding to the lowest cost value. This solution serves as initial guess for the regularized optimization problem.

\begin{figure}[!t]
	\begin{center}
		\includegraphics[width=0.9\textwidth]{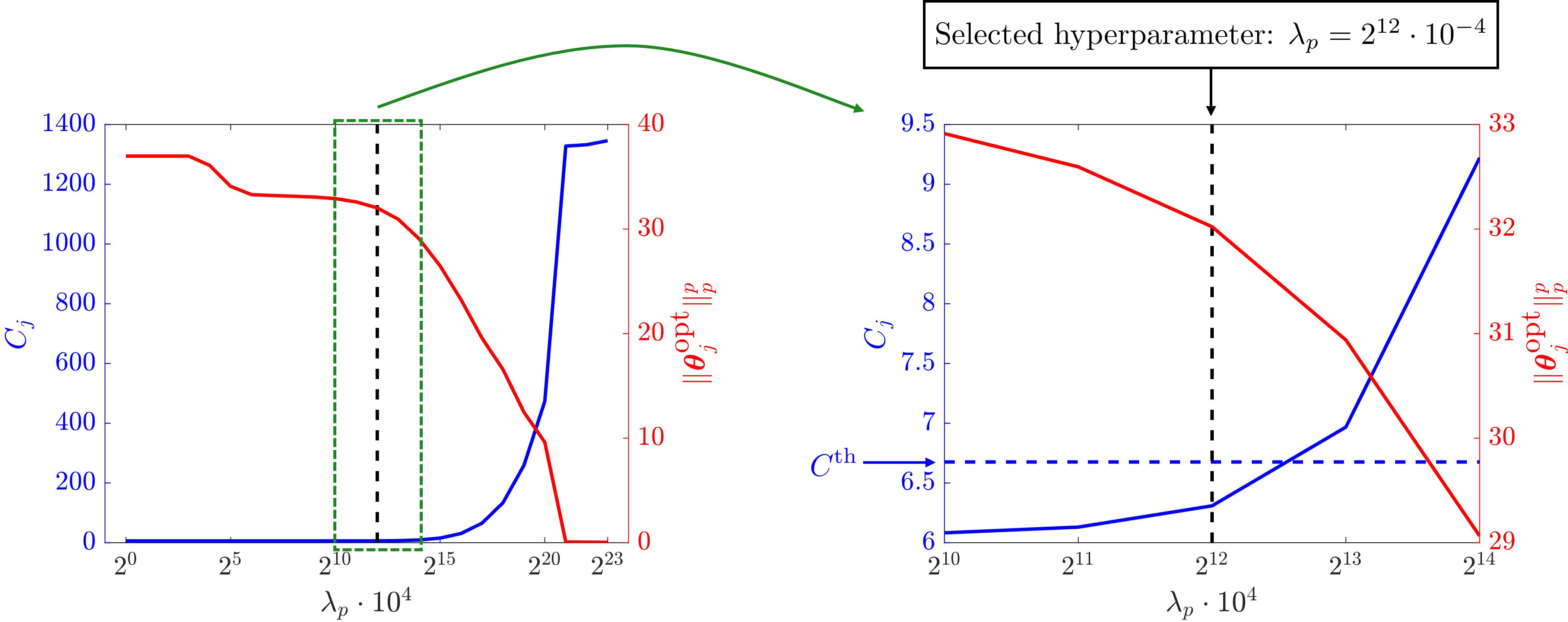}
		\caption{Pareto analysis for the automated selection of the hyperparameter $\lambda_p$. The left figure shows the cost function values (in $\text{kN}^2$) and the regularization term values as functions of the different choices of $\lambda_p$ (in $\text{kN}^2$). A magnification in the region around the automatically selected hyperparameter is shown in the right figure. The threshold for the cost function value and the solution obtained from the automated selection algorithm are highlighted with blue and black dashed lines, respectively. The solutions were obtained from the data corresponding to model LEVP with noise level $\sigma = 0.5 \mu \text{m}$ (see \sectionname\ref{sec:benchmarks})}
		\label{fig:Pareto}
	\end{center}
\end{figure}

The value of the hyperparameter $\lambda_p$ is of paramount importance for the achievement of a good compromise between accuracy and sparsity of the discovered material model. As follows, we propose an automated algorithm for its selection. We solve \eqref{eq:objective_regularized} for multiple different values of $\lambda_p$ which are successively increased by a factor of two, starting from an initial small value. In this work, we adopt $\lambda_p \in \{10^{-4} \cdot 2^j \ \text{kN}^2 : j=0,\dots,23\}$. For every value of $\lambda_p$, \eqref{eq:objective_regularized} delivers a solution for the material parameters and an associated value of the cost function, which we store in the set
\be
\mathcal{S} = \{ (\bftheta^{\text{opt}}_j,C_j) : j=1,\dots,n_\lambda \},
\ee
with $n_\lambda = 24$ as the number of considered $\lambda_p$ values. 
\figurename\ref{fig:Pareto} shows an exemplary plot of the cost function and regularization term values for the different choices of $\lambda_p$, which follows a consistently encountered trend. As $\lambda_p$ increases, the cost function value increases and the regularization term value decreases, implying that the discovered model becomes decreasingly accurate and increasingly simple. It is thus evident that this plot carries the information needed to strike a balance between the conflicting objectives of accuracy and parsimony. Importantly, the initial range of $\lambda_p$ values is characterized by a low rate of variation of both cost and regularization term, so that at least visually it is not difficult to identify a threshold value of $\lambda_p$ beyond which the cost start increasing (and the regularization term decreasing) at a much faster rate. Rather than defining a threshold on $\lambda_p$, we prefer to define one on the cost, $C^{\text{th}}$, with the objective to limit the tolerated error on the fitting accuracy of the discovered models. In this work, we choose $C^{\text{th}}$ to be slightly larger than the smallest cost $C^\text{min}$ in the solution set $\mathcal{S}$ but not smaller than $10^{-5} \ \text{kN}^2$, i.e., $C^{\text{th}} = \max \{ 10^{-5} \ \text{kN}^2 , 1.1 \ C^{\text{min}} \}$. Once this threshold is chosen, all solutions with $C_j \geq C^{\text{th}}$ are considered irrelevant. The remaining relevant solutions are
\be
\mathcal{S}^{\text{th}} = \{ (\bftheta^{\text{opt}}_j,C_j)\in\mathcal{S} : C_j < C^{\text{th}} \},
\ee
from which we select the sparsest solution, i.e., the solution in $\mathcal{S}^{\text{th}}$ with the smallest value of $\|\bftheta^{\text{opt}}_j\|_p^p$. In all cases, this automated selection algorithm leads to a solution with both low cost and low regularization term (see \figurename\ref{fig:Pareto}), thus striking a user-defined compromise between model accuracy and sparsity.



\subsection{Material classification}
In a final postprocessing step, after having selected a suitable value for $\lambda_p$, we set all components of $\bftheta$ that lie below a threshold $\theta^{\text{th}}$ (chosen as $\theta^{\text{th}}=10^{-4}$ in this work) to their minimal value, i.e., to the corresponding value in $\bftheta^{\text{min}}$.
Further, if the product of the modulus and the reciprocal of a relaxation time of a Maxwell element lies below the threshold, both the modulus and the reciprocal of the relaxation time are set to their minimal values.
In this way, unnecessary terms in the thermodynamic potentials are removed and the discovered material can be classified into different material classes. If $G$ or $K$ are larger than the threshold, the material is classified as elastic. If one of the reciprocal relaxation times $1/g_i$ or $1/k_i$ is larger than the threshold, the material is classified as viscoelastic. If, instead, any of the reciprocal relaxation times $1/g_i$ or $1/k_i$ is below the threshold, we assume $g_i \rightarrow \infty$ or $k_i \rightarrow \infty$, respectively. If the reciprocal of the yield stress is below the threshold, neither plasticity nor viscoplasticity is active, and we assume $\sigma_0 \rightarrow \infty$. If, however, the reciprocal of the yield stress is larger than the threshold, the material is classified as plastic and the value of $\eta_p$ decides whether the material is further classified as viscoplastic. Finally, isotropic or kinematic hardening are active if respectively $H^{iso}$ or $H^{kin}$ are larger than the threshold.


\section{Numerical benchmarks}
\label{sec:benchmarks}

To assess the performance of the developed method, we test it on synthetic data generated by finite element simulations. While testing on real experimental data will be an important future objective, the deployment of computational data has the obvious advantage that the ground truth, i.e., the true material model underlying the data, is exactly known. 
We select five different benchmark material models:
\begin{itemize}
\itemsep0em
\item {E}: Elastic
\item {VE}: Viscoelastic
\item {VEEP}: Viscoelastic (only shear) \& elastoplastic with isotropic hardening 
\item {EVP}: Elastic \& viscoplastic with kinematic hardening 
\item {VEVP}: Viscoelastic \& viscoplastic with mixed hardening 
\end{itemize}

\tablename\ref{tab:mat_types} illustrates more in detail the features active within each benchmark material model, while the corresponding material parameters are given in \tablename\ref{tab:mat_parameters}.
\subsection{Data generation}
\label{sec:data_generation}

\begin{figure}[!t]
	\begin{center}
		\includegraphics[width=0.75\textwidth]{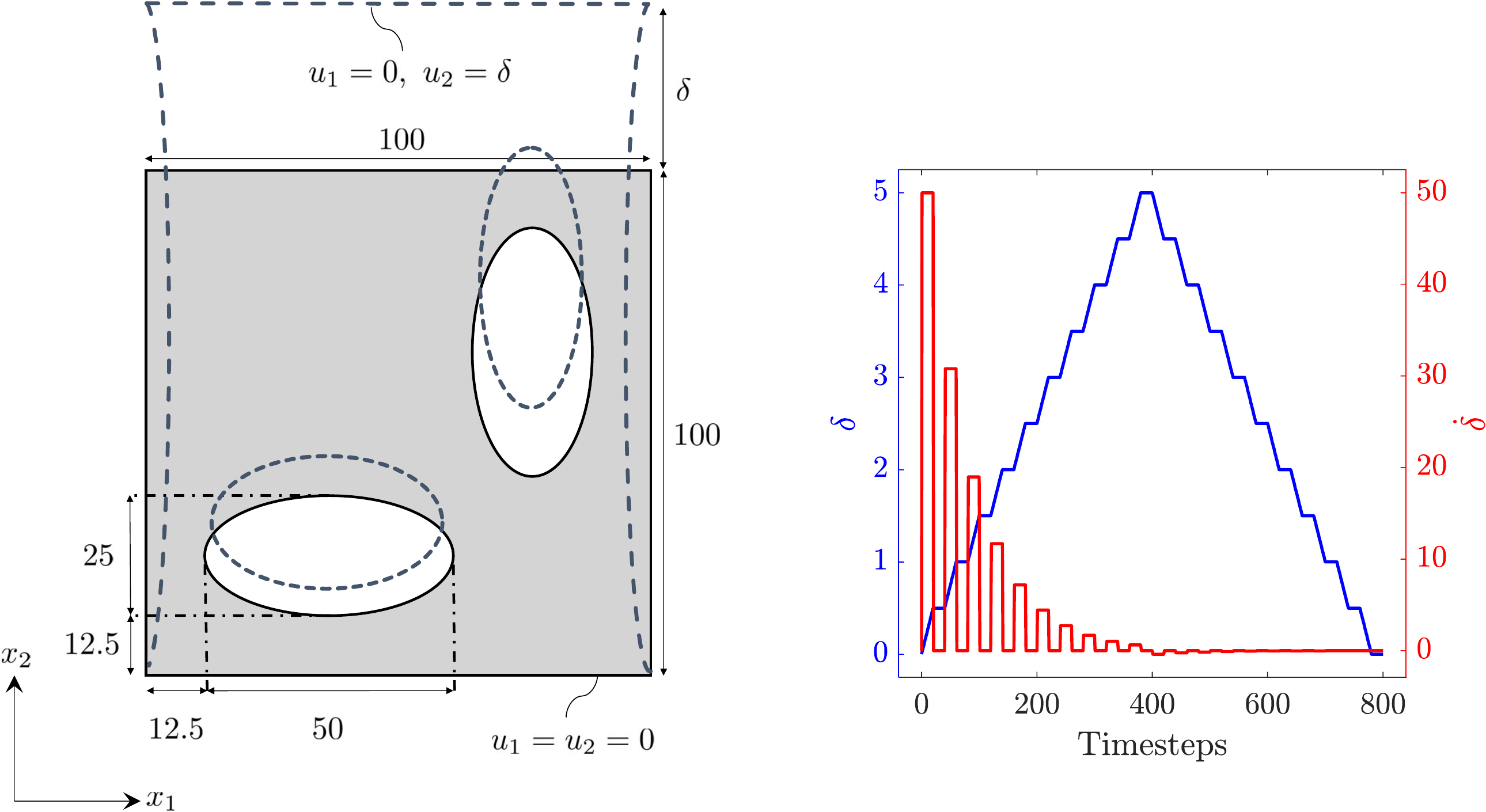}
		\caption{Displacement-controlled loading conditions of a square domain with two elliptic holes. The dimensions of the specimen and the loading parameter $\delta$ are in $\text{mm}$ and the loading rate $\dot\delta$ is in $\text{mm}/\text{s}$.}
		\label{fig:bcs}
	\end{center}
\end{figure}

For each benchmark material model, a two-dimensional finite element simulation under plane-strain conditions is conducted. To obtain a heterogeneous strain field and hence a rich data set during the experiment, complex geometries are preferred over simple geometries. We therefore adopt a square specimen with two elliptic holes as depicted in \figurename\ref{fig:bcs} (left), which has been already shown to produce a diverse strain field upon deformation (see \cite{flaschel_discovering_2022}), and we load it under displacement-controlled tension and compression. The loading parameter $\delta$ determines the displacement at the upper boundary of the specimen. As the model library considered in this work includes a few rate-dependent models, different rates of the loading parameter $\dot\delta$ are included during data generation as shown in \figurename\ref{fig:bcs} (right). In particular, we consider a number of $n_{\idxPHASE}=20$ loading phases, indexed by ${\idxPHASE}=1,\dots,n_{\idxPHASE}$, out of which $n_{\idxPHASE}/2$ under tension and $n_{\idxPHASE}/2$ under compression. Each loading phase consists of a loading period, during which $\dot\delta$ is constant, and a subsequent relaxation period, during which no further deformation is applied at the boundary ($\dot\delta=0$). During each loading period, the same absolute value of deformation $|\delta| = 0.5 \text{mm}$ is applied over different loading time intervals $t^{\text{load}}_{\idxPHASE}$, which are chosen evenly spaced on a logarithmic scale between $t^{\text{load}}_1 = 10^{-2} \text{s}$ and $t^{\text{load}}_{20} = 10^{2} \text{s}$, i.e., $t^{\text{load}}_{\idxPHASE} \in \{ 10^{ ( -2 + \frac{4}{n_{\idxPHASE}-1} i ) } \text{s} : i=0,\dots,(n_k-1)\}$. The time intervals of the relaxation periods are assumed all equal to $t^{\text{relax}}_{\idxPHASE}=10^{2} \text{s}$. The entire problem is discretized in time with a total number of $n_t = 800$ time steps, choosing the same number of time steps for each loading and relaxation period.

From the finite element simulations, the nodal displacements and net reaction force at the upper specimen boundary are extracted. To mimic real experiments, we add  to the data artificial Gaussian noise as follows
\be
u^{a,\idxLOAD}_i = u^{\text{fem},a,\idxLOAD}_i + u^{\text{noise},a,\idxLOAD}_i, \quad u^{\text{noise},a,\idxLOAD}_i \sim \calN(0,\sigma) \quad \forall \quad a\in\{1,\dots,n_n\},~i\in\{1,2\},~\idxLOAD\in\{1,\dots,n_t\},
\ee
where $u^{\text{fem},a,\idxLOAD}_i$ is the $i^{\text{th}}$ displacement component of node $a$ at the $t^\text{th}$ time step obtained from the finite element simulations, and $u^{\text{noise},a,\idxLOAD}_i$ denotes the added noise, sampled from a Gaussian distribution with zero mean and standard deviation $\sigma$. Considering modern DIC setups, a reasonable upper limit for the noise is $\sigma=0.1~\mu\text{m}$ \citep{pierron_extension_2010,marek_extension_2019}. For our benchmarks, we test different noise levels $\sigma \in \{0 \mu\text{m},0.1 \mu \text{m},0.3 \mu \text{m},0.5 \mu \text{m}\}$. After applying artificial noise to the data, temporal Savitzky–Golay denoising \citep{savitzky_smoothing_1964} with quadratic polynomial order and a moving-window length of $10$ 
time steps is applied to each loading period and relaxation period. 

\subsection{Discovered material models}
EUCLID is applied to the full-field displacement and reaction force data obtained from the five different benchmark models with different noise levels to discover the hidden material behavior. \tablename\ref{tab:mat_types} illustrates the discovered material classes in comparison to the true ones, revealing that EUCLID is indeed able to automatically discriminate among different classes of material behavior. In other words, purely based on displacement and force data on one single test, EUCLID is able to discern whether the hidden material behavior is elastic, viscoelastic, plastic or viscoplastic, and whether isotropic or kinematic hardening takes place. 
Note that the automatic selection of the few terms in the thermodynamic potentials needed to describe the material behavior implies the automatic selection of the correspondingly active internal variables.
Only for one benchmark, namely, the VEEP model  with noise level $\sigma = 0.3 \mu \text{m}$, EUCLID fails to predict the true material class as it discovers a Maxwell element for the bulk material response, which is not contained in the true material model. The discovered material model is nevertheless parsimonious, as the effects of rate-dependent plasticity and kinematic hardening are correctly excluded from the material response.

\begin{table*}[t]
\centering
\caption{True and discovered material types for different noise levels $\sigma$ (in $\text{mm}$).}
\label{tab:mat_types}
\begin{tabular}{llcccccccc}
\multicolumn{2}{c}{Benchmarks} &
\rotatebox{90}{Elasticity} &
\rotatebox{90}{Viscoelasticity} &
\rotatebox{90}{\# Maxwell (shear)} &
\rotatebox{90}{\# Maxwell (bulk)} &
\rotatebox{90}{Plasticity} &
\rotatebox{90}{Viscoplasticity} &
\rotatebox{90}{Isotropic Hardening} &
\rotatebox{90}{Kinematic Hardening} \\%
\hline\\[-10pt]
E & Truth & \cellcolor{green!15} on & \cellcolor{red!15} off & \cellcolor{red!15} 0 & \cellcolor{red!15} 0 & \cellcolor{red!15} off & \cellcolor{red!15} off & \cellcolor{red!15} off & \cellcolor{red!15} off \\%
~ & \CA $\sigma=0 $& \cellcolor{green!15} on & \cellcolor{red!15} off & \cellcolor{red!15} 0 & \cellcolor{red!15} 0 & \cellcolor{red!15} off & \cellcolor{red!15} off & \cellcolor{red!15} off & \cellcolor{red!15} off \\%
~ & \CB $\sigma=10^{-4}$ & \cellcolor{green!15} on & \cellcolor{red!15} off & \cellcolor{red!15} 0 & \cellcolor{red!15} 0 & \cellcolor{red!15} off & \cellcolor{red!15} off & \cellcolor{red!15} off & \cellcolor{red!15} off \\%
~ & \CC $\sigma=3 \cdot 10^{-4}$ & \cellcolor{green!15} on & \cellcolor{red!15} off & \cellcolor{red!15} 0 & \cellcolor{red!15} 0 & \cellcolor{red!15} off & \cellcolor{red!15} off & \cellcolor{red!15} off & \cellcolor{red!15} off \\%
~ & \CD $\sigma=5 \cdot 10^{-4}$ & \cellcolor{green!15} on & \cellcolor{red!15} off & \cellcolor{red!15} 0 & \cellcolor{red!15} 0 & \cellcolor{red!15} off & \cellcolor{red!15} off & \cellcolor{red!15} off & \cellcolor{red!15} off \\%
\hline\\[-10pt]
VE & Truth & \cellcolor{green!15} on & \cellcolor{green!15} on & \cellcolor{green!15} 1 & \cellcolor{green!15} 1 & \cellcolor{red!15} off & \cellcolor{red!15} off & \cellcolor{red!15} off & \cellcolor{red!15} off \\%
~ & \CA $\sigma=0 $& \cellcolor{green!15} on & \cellcolor{green!15} on & \cellcolor{green!15} 1 & \cellcolor{green!15} 1 & \cellcolor{red!15} off & \cellcolor{red!15} off & \cellcolor{red!15} off & \cellcolor{red!15} off \\%
~ & \CB $\sigma=10^{-4}$ & \cellcolor{green!15} on & \cellcolor{green!15} on & \cellcolor{green!15} 1 & \cellcolor{green!15} 1 & \cellcolor{red!15} off & \cellcolor{red!15} off & \cellcolor{red!15} off & \cellcolor{red!15} off \\%
~ & \CC $\sigma=3 \cdot 10^{-4}$ & \cellcolor{green!15} on & \cellcolor{green!15} on & \cellcolor{green!15} 1 & \cellcolor{green!15} 1 & \cellcolor{red!15} off & \cellcolor{red!15} off & \cellcolor{red!15} off & \cellcolor{red!15} off \\%
~ & \CD $\sigma=5 \cdot 10^{-4}$ & \cellcolor{green!15} on & \cellcolor{green!15} on & \cellcolor{green!15} 1 & \cellcolor{green!15} 1 & \cellcolor{red!15} off & \cellcolor{red!15} off & \cellcolor{red!15} off & \cellcolor{red!15} off \\%
\hline\\[-10pt]
VEEP & Truth & \cellcolor{green!15} on & \cellcolor{green!15} on & \cellcolor{green!15} 1 & \cellcolor{red!15} 0 & \cellcolor{green!15} on & \cellcolor{red!15} off & \cellcolor{green!15} on & \cellcolor{red!15} off \\%
~ & \CA $\sigma=0 $& \cellcolor{green!15} on & \cellcolor{green!15} on & \cellcolor{green!15} 1 & \cellcolor{red!15} 0 & \cellcolor{green!15} on & \cellcolor{red!15} off & \cellcolor{green!15} on & \cellcolor{red!15} off \\%
~ & \CB $\sigma=10^{-4}$ & \cellcolor{green!15} on & \cellcolor{green!15} on & \cellcolor{green!15} 1 & \cellcolor{red!15} 0 & \cellcolor{green!15} on & \cellcolor{red!15} off & \cellcolor{green!15} on & \cellcolor{red!15} off \\%
~ & \CC $\sigma=3 \cdot 10^{-4}$ & \cellcolor{green!15} on & \cellcolor{green!15} on & \cellcolor{green!15} 1 & \cellcolor{green!15} 1 & \cellcolor{green!15} on & \cellcolor{red!15} off & \cellcolor{green!15} on & \cellcolor{red!15} off \\%
~ & \CD $\sigma=5 \cdot 10^{-4}$ & \cellcolor{green!15} on & \cellcolor{green!15} on & \cellcolor{green!15} 1 & \cellcolor{red!15} 0 & \cellcolor{green!15} on & \cellcolor{red!15} off & \cellcolor{green!15} on & \cellcolor{red!15} off \\%
\hline\\[-10pt]
EVP & Truth & \cellcolor{green!15} on & \cellcolor{red!15} off & \cellcolor{red!15} 0 & \cellcolor{red!15} 0 & \cellcolor{green!15} on & \cellcolor{green!15} on & \cellcolor{red!15} off & \cellcolor{green!15} on \\%
~ & \CA $\sigma=0 $& \cellcolor{green!15} on & \cellcolor{red!15} off & \cellcolor{red!15} 0 & \cellcolor{red!15} 0 & \cellcolor{green!15} on & \cellcolor{green!15} on & \cellcolor{red!15} off & \cellcolor{green!15} on \\%
~ & \CB $\sigma=10^{-4}$ & \cellcolor{green!15} on & \cellcolor{red!15} off & \cellcolor{red!15} 0 & \cellcolor{red!15} 0 & \cellcolor{green!15} on & \cellcolor{green!15} on & \cellcolor{red!15} off & \cellcolor{green!15} on \\%
~ & \CC $\sigma=3 \cdot 10^{-4}$ & \cellcolor{green!15} on & \cellcolor{red!15} off & \cellcolor{red!15} 0 & \cellcolor{red!15} 0 & \cellcolor{green!15} on & \cellcolor{green!15} on & \cellcolor{red!15} off & \cellcolor{green!15} on \\%
~ & \CD $\sigma=5 \cdot 10^{-4}$ & \cellcolor{green!15} on & \cellcolor{red!15} off & \cellcolor{red!15} 0 & \cellcolor{red!15} 0 & \cellcolor{green!15} on & \cellcolor{green!15} on & \cellcolor{red!15} off & \cellcolor{green!15} on \\%
\hline\\[-10pt]
VEVP & Truth & \cellcolor{green!15} on & \cellcolor{green!15} on & \cellcolor{green!15} 1 & \cellcolor{green!15} 1 & \cellcolor{green!15} on & \cellcolor{green!15} on & \cellcolor{green!15} on & \cellcolor{green!15} on \\%
~ & \CA $\sigma=0 $& \cellcolor{green!15} on & \cellcolor{green!15} on & \cellcolor{green!15} 1 & \cellcolor{green!15} 1 & \cellcolor{green!15} on & \cellcolor{green!15} on & \cellcolor{green!15} on & \cellcolor{green!15} on \\%
~ & \CB $\sigma=10^{-4}$ & \cellcolor{green!15} on & \cellcolor{green!15} on & \cellcolor{green!15} 1 & \cellcolor{green!15} 1 & \cellcolor{green!15} on & \cellcolor{green!15} on & \cellcolor{green!15} on & \cellcolor{green!15} on \\%
~ & \CC $\sigma=3 \cdot 10^{-4}$ & \cellcolor{green!15} on & \cellcolor{green!15} on & \cellcolor{green!15} 1 & \cellcolor{green!15} 1 & \cellcolor{green!15} on & \cellcolor{green!15} on & \cellcolor{green!15} on & \cellcolor{green!15} on \\%
~ & \CD $\sigma=5 \cdot 10^{-4}$ & \cellcolor{green!15} on & \cellcolor{green!15} on & \cellcolor{green!15} 1 & \cellcolor{green!15} 1 & \cellcolor{green!15} on & \cellcolor{green!15} on & \cellcolor{green!15} on & \cellcolor{green!15} on \\%
\hline\\[-10pt]
\end{tabular}%
\end{table*}

Besides and simultaneously to identifying the material class, EUCLID computes the unknown material parameters. \tablename\ref{tab:mat_parameters} reports the comparison between the true and the calibrated material parameters for the different benchmarks, revealing a very good agreement. As expected, the discrepancy between true and discovered parameters increases for increasing noise level.

\begin{table*}[t]
\centering
\caption{True and discovered material parameters for different noise levels $\sigma$ (in $\text{mm}$).}
\label{tab:mat_parameters}
\resizebox{16cm}{!}{
\begin{tabular}{llllllllllll}
\multicolumn{2}{c}{Benchmarks} & $G$ $\left[ \frac{\text{kN}}{\text{mm}^2} \right]$ & $K$ $\left[ \frac{\text{kN}}{\text{mm}^2} \right]$ & $G_1$ $\left[ \frac{\text{kN}}{\text{mm}^2} \right]$ & $g_1$ $\left[ \text{s} \right]$ & $K_1$ $\left[ \frac{\text{kN}}{\text{mm}^2} \right]$ & $k_1$ $\left[ \text{s} \right]$ & $\sigma_0$ $\left[ \frac{\text{kN}}{\text{mm}^2} \right]$ & $\eta_p$ $\left[ \frac{\text{kN} \ \text{s}}{\text{mm}^2} \right]$ & $H^{iso}$ $\left[ \frac{\text{kN}}{\text{mm}^2} \right]$ & $H^{kin}$ $\left[ \frac{\text{kN}}{\text{mm}^2} \right]$ \\[5pt]%
\hline\\[-10pt]
E & Truth & 0.6000 & 1.3000 & 0 & $\rightarrow\infty$ & 0 & $\rightarrow\infty$ & $\rightarrow\infty$ & 0 & 0 & 0 \\%
~ & \CA $\sigma=0 $& 0.6000 & 1.3000 & 0 & $\rightarrow\infty$ & 0 & $\rightarrow\infty$ & $\rightarrow\infty$ & 0 & 0 & 0 \\%
~ & \CB $\sigma=10^{-4}$ & 0.6000 & 1.3000 & 0 & $\rightarrow\infty$ & 0 & $\rightarrow\infty$ & $\rightarrow\infty$ & 0 & 0 & 0 \\%
~ & \CC $\sigma=3 \cdot 10^{-4}$ & 0.5998 & 1.3004 & 0 & $\rightarrow\infty$ & 0 & $\rightarrow\infty$ & $\rightarrow\infty$ & 0 & 0 & 0 \\%
~ & \CD $\sigma=5 \cdot 10^{-4}$ & 0.5998 & 1.3002 & 0 & $\rightarrow\infty$ & 0 & $\rightarrow\infty$ & $\rightarrow\infty$ & 0 & 0 & 0 \\%
\hline\\[-10pt]
VE & Truth & 0.6000 & 1.3000 & 0.3500 & 110.0000 & 0.4000 & 15.0000 & $\rightarrow\infty$ & 0 & 0 & 0 \\%
~ & \CA $\sigma=0 $& 0.6000 & 1.2999 & 0.3500 & 109.9953 & 0.3999 & 15.0047 & $\rightarrow\infty$ & 0 & 0 & 0 \\%
~ & \CB $\sigma=10^{-4}$ & 0.6017 & 1.2930 & 0.3517 & 109.6126 & 0.3929 & 15.5063 & $\rightarrow\infty$ & 0 & 0 & 0 \\%
~ & \CC $\sigma=3 \cdot 10^{-4}$ & 0.6042 & 1.2674 & 0.3544 & 109.7273 & 0.3671 & 16.5166 & $\rightarrow\infty$ & 0 & 0 & 0 \\%
~ & \CD $\sigma=5 \cdot 10^{-4}$ & 0.6094 & 1.2702 & 0.3597 & 108.2126 & 0.3693 & 15.1302 & $\rightarrow\infty$ & 0 & 0 & 0 \\%
\hline\\[-10pt]
VEEP & Truth & 0.6000 & 1.3000 & 0.3500 & 110.0000 & 0 & $\rightarrow\infty$ & 0.0300 & 0 & 0.0300 & 0 \\%
~ & \CA $\sigma=0 $& 0.6000 & 1.3000 & 0.3500 & 110.0008 & 0 & $\rightarrow\infty$ & 0.0300 & 0 & 0.0300 & 0 \\%
~ & \CB $\sigma=10^{-4}$ & 0.6008 & 1.2997 & 0.3506 & 109.5033 & 0 & $\rightarrow\infty$ & 0.0303 & 0 & 0.0276 & 0 \\%
~ & \CC $\sigma=3 \cdot 10^{-4}$ & 0.5935 & 1.3055 & 0.3435 & 114.3310 & 0.0090 & 31.5015 & 0.0314 & 0 & 0.0213 & 0 \\%
~ & \CD $\sigma=5 \cdot 10^{-4}$ & 0.5908 & 1.2963 & 0.3409 & 116.6722 & 0 & $\rightarrow\infty$ & 0.0329 & 0 & 0.0123 & 0 \\%
\hline\\[-10pt]
EVP & Truth & 0.6000 & 1.3000 & 0 & $\rightarrow\infty$ & 0 & $\rightarrow\infty$ & 0.0300 & 0.0400 & 0 & 0.0100 \\%
~ & \CA $\sigma=0 $& 0.6000 & 1.3000 & 0 & $\rightarrow\infty$ & 0 & $\rightarrow\infty$ & 0.0300 & 0.0399 & 0 & 0.0100 \\%
~ & \CB $\sigma=10^{-4}$ & 0.6013 & 1.2997 & 0 & $\rightarrow\infty$ & 0 & $\rightarrow\infty$ & 0.0303 & 0.0360 & 0 & 0.0099 \\%
~ & \CC $\sigma=3 \cdot 10^{-4}$ & 0.6015 & 1.2991 & 0 & $\rightarrow\infty$ & 0 & $\rightarrow\infty$ & 0.0307 & 0.0288 & 0 & 0.0099 \\%
~ & \CD $\sigma=5 \cdot 10^{-4}$ & 0.6020 & 1.2965 & 0 & $\rightarrow\infty$ & 0 & $\rightarrow\infty$ & 0.0313 & 0.0229 & 0 & 0.0096 \\%
\hline\\[-10pt]
VEVP & Truth & 0.6000 & 1.3000 & 0.3500 & 110.0000 & 0.4000 & 15.0000 & 0.0300 & 0.0400 & 0.0300 & 0.0100 \\%
~ & \CA $\sigma=0 $& 0.6000 & 1.3000 & 0.3500 & 110.0096 & 0.4000 & 15.0007 & 0.0300 & 0.0399 & 0.0300 & 0.0100 \\%
~ & \CB $\sigma=10^{-4}$ & 0.5996 & 1.3011 & 0.3494 & 110.5654 & 0.4020 & 15.1755 & 0.0304 & 0.0321 & 0.0276 & 0.0098 \\%
~ & \CC $\sigma=3 \cdot 10^{-4}$ & 0.5970 & 1.2930 & 0.3466 & 113.9556 & 0.3965 & 14.6939 & 0.0316 & 0.0158 & 0.0201 & 0.0095 \\%
~ & \CD $\sigma=5 \cdot 10^{-4}$ & 0.5967 & 1.3001 & 0.3468 & 114.2066 & 0.4026 & 14.7275 & 0.0320 & 0.0160 & 0.0181 & 0.0094 \\%
\hline\\[-10pt]
\end{tabular}%
}%
\end{table*}

\begin{figure}[!t]
	\begin{center}
		\includegraphics[width=0.99\textwidth]{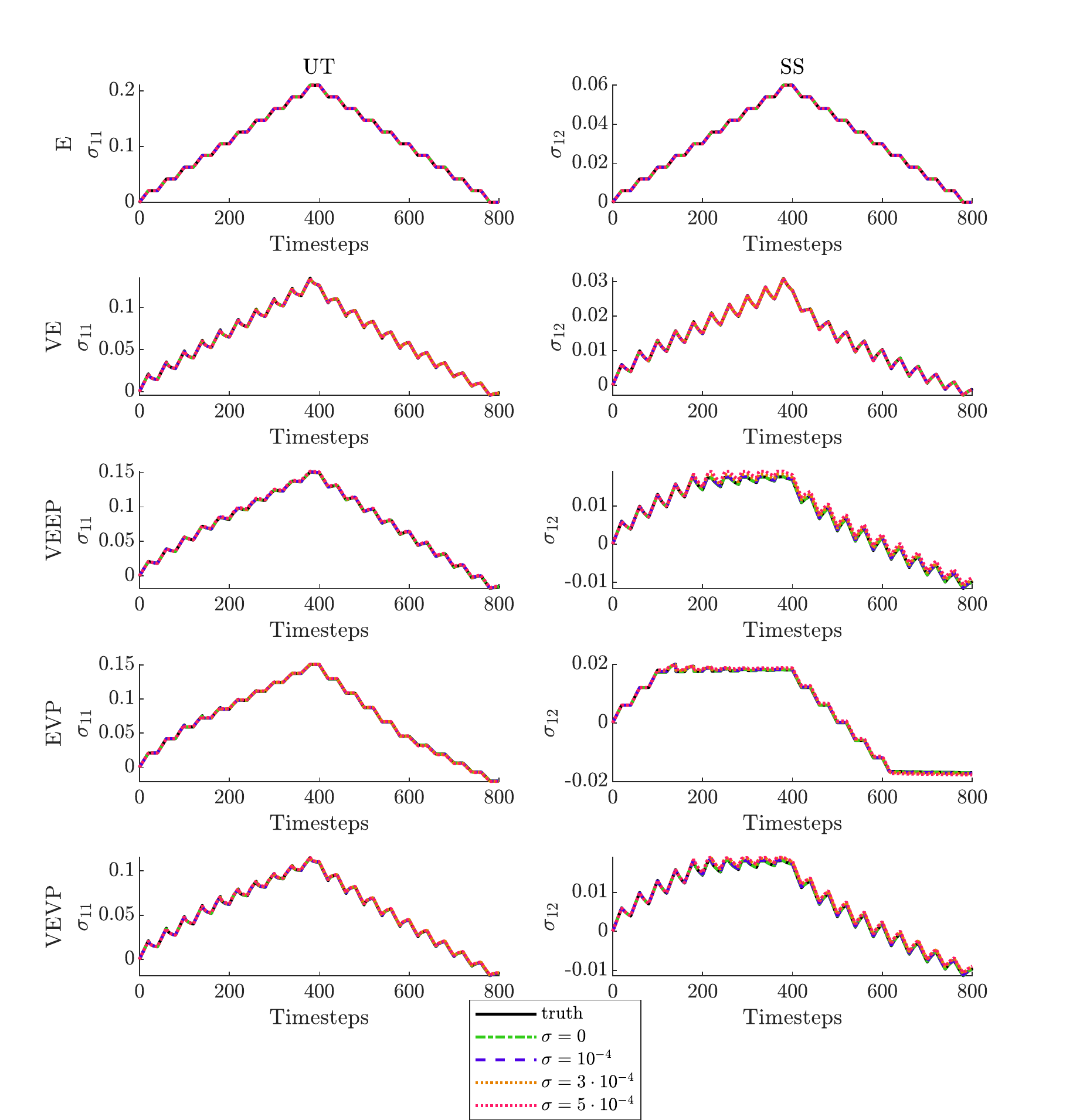}
		\caption{True and discovered material responses under uniaxial tension (UT) and simple shear (SS) for different noise levels $\sigma$ (in $\text{mm}$). The stress components $\sigma_{11}$ and $\sigma_{12}$ are in $\frac{\text{kN}}{\text{mm}^2}$.}
		\label{fig:load_path}
	\end{center}
\end{figure}

Finally, in order to test if the discovered material models describe a similar material response as the true ones under unseen loading histories (and especially for the cases with noise, featuring some discrepancy between true and discovered material parameters), we compute the stress response of the material along two different arbitrarily chosen deformation paths, see \figurename\ref{fig:load_path}. We select these strain paths as uniaxial tension (UT) and simple shear (SS), given by
\be
\bfeps^{\text{UT}} = 
\begin{bmatrix}
\epsilon & 0 & 0\\
0 & 0 & 0\\
0 & 0 & 0\\
\end{bmatrix}, \qquad
\bfeps^{\text{SS}} = 
\begin{bmatrix}
0 & \epsilon & 0\\
\epsilon & 0 & 0\\
0 & 0 & 0\\
\end{bmatrix},
\ee
where $\epsilon$ is a scalar deformation parameter that is let to increase to a value of $0.1$ during ten loading phases and to subsequently decrease to zero during additional ten loading phases. The loading and relaxation periods of each loading phase, as well as the loading and relaxation times, are chosen in the same way as described for the loading parameter $\delta$ in \sectionname\ref{sec:data_generation}.
\figurename\ref{fig:load_path} demonstrates the good agreement between true and discovered material responses, with discrepancies that, as expected, increase for increasing noise level.

\section{Conclusions}\label{sec:conclusions}
We extended the scope of our recently developed EUCLID approach for unsupervised automated discovery of material laws to the highly general case of a material belonging to an unknown class of constitutive behavior. To this end, we leveraged the theory of generalized standard materials, which encompasses a large number of important constitutive classes including elasticity, viscosity, plasticity and arbitrary combinations thereof. We showed that, based solely on full-field displacements and net reaction forces, EUCLID is able to automatically discover the two scalar thermodynamic potentials, namely, the Helmholtz free energy and the dissipation potential, which completely define the behavior of generalized standard materials. Stability and thermodynamic consistency of the discovered model are guaranteed by construction by the a priori requirement of convexity for the two potentials; balance of linear momentum (enforced locally in the interior and globally at each constrained side of the domain) acts as a fundamental constraint to counteract the ill-posedness of the inverse problem due to the unavailability of stress-strain labeled pairs; sparsity promoting regularization leads to the automatic selection of a small subset from a possibly large number of candidate model features in the initially chosen large model library and thus leads to a parsimonious, i.e., simple and interpretable, model. Importantly, since model features go hand in hand with the corresponding internal variables, sparse regression automatically induces a parsimonious selection of the few internal variables needed for an accurate but simple description of the material behavior. We propose a fully automatic procedure for the selection of the hyperparameter controlling the weight of the sparsity promoting regularization term, in order to strike a user-defined balance between model accuracy and simplicity.

The present study paves the way for a number of interesting extensions. A possible important direction would be a further generalization of the model library for the thermodynamic potentials to include additional phenomena such as anisotropy, heterogeneity, softening (which would require the consideration of gradient terms), coupled thermo- or electromechanical effects, to name but a few.
Another important aim of future research would be the enhancement of the computational efficiency, as solving the constitutive equations becomes more demanding for an increasing number of unknown parameters in the inverse problem.
To ease the implementation of the constitutive equation solver for a wide library of thermodynamic potentials, relying on automatic differentiation would be beneficial; however,  as discussed in this paper, automatic differentiation requires a regularization of the potentials and currently implies a decrease in the overall computational efficiency, which could be addressed in the future. Last but not least, testing of EUCLID on experimental data would be an important future goal.

\section*{Acknowledgments}

MF and LDL would like to acknowledge funding by SNF through grant N. $200021\_204316$ ``Unsupervised data-driven discovery of material laws''.

\appendix

\section{Solving the constitutive equations}
\label{sec:stress_update}
We solve the constitutive equations (see \sectionname\ref{sec:model_library_constitutive_laws}) in strain control. Given the state variables and dependent variables at the previous time step $(t-1)$ and the strain $\bfepsilon^t$ at the current time step $t$, the objective is to compute first the internal variables $\bfalpha^t$ using \eqnames\eqref{eq:evolution_VE},~\eqref{eq:evolution_VP} and then the stress $\bfsigma^t$ using \eqname\eqref{eq:stress}. To this end, \eqnames\eqref{eq:evolution_VE},~\eqref{eq:evolution_VP},~\eqref{eq:stress} are discretized in time using an implicit Euler scheme, i.e., $\dot\square^t = \frac{1}{\Delta t}\left(\square^t-\square^{t-1}\right)$, where $\Delta\square = \square^t-\square^{t-1}$ is used to denote differences of variables between the previous and current time step. The discretized equations are then solved with a viscoelastic predictor / viscoplastic corrector algorithm (see \cite{simo_computational_1998,neto_computational_2008} for similar algorithms).

\subsection{Viscoelastic predictor step}
\label{sec:viscoelastic_predictor}
For the viscoelastic predictor step (also known as trial step), we assume a purely viscoelastic behavior, i.e., $f^{t,\text{trial}} \leq 0$. Under this assumption, the viscoplastic internal variables in \eqnames\eqref{eq:evolution_VP} do not evolve, e.g., $\boldsymbol{\alpha}_{\RomanNumeralCaps{1}}^{t,\text{trial}}=\boldsymbol{\alpha}_{\RomanNumeralCaps{1}}^{t-1}$. The viscoelastic internal variables are updated using the time-discretized \eqname\eqref{eq:evolution_VE}
\begin{equation}
\frac{1}{\Delta t}\left(\boldsymbol{\alpha}_j^{t,\text{trial}} - \boldsymbol{\alpha}_j^{t-1}\right)
= \frac{1}{g_j} \dev(\bfepsilon^t - \boldsymbol{\alpha}_j^{t,\text{trial}} - \boldsymbol{\alpha}_{\RomanNumeralCaps{1}}^{t-1}) + \frac{1}{k_j} \vol(\bfepsilon^t - \boldsymbol{\alpha}_j^{t,\text{trial}} - \boldsymbol{\alpha}_{\RomanNumeralCaps{1}}^{t-1}).
\end{equation}
Splitting this equation into its deviatoric and volumetric contributions, we obtain explicit expressions for the update of the viscoelastic internal variables
\begin{equation}
\begin{aligned}
\dev\boldsymbol{\alpha}_j^{t,\text{trial}} &= \frac{1}{\left( 1+\frac{\Delta t}{g_j} \right)} \left( \dev\boldsymbol{\alpha}_j^{t-1} + \frac{\Delta t}{g_j} \dev\left(\bfepsilon^t-\boldsymbol{\alpha}_{\RomanNumeralCaps{1}}^{t-1}\right) \right) ,\\
\vol\boldsymbol{\alpha}_j^{t,\text{trial}} &= \frac{1}{\left( 1+\frac{\Delta t}{k_j} \right)} \left( \vol\boldsymbol{\alpha}_j^{t-1} + \frac{\Delta t}{k_j} \vol\left(\bfepsilon^t-\boldsymbol{\alpha}_{\RomanNumeralCaps{1}}^{t-1}\right) \right).\\
\end{aligned}
\end{equation}
Knowing all internal variables at the current time step $t$, the trial stress $\bfsigma^{t,\text{trial}}$ is computed using \eqname\eqref{eq:stress}.
After some algebra, we arrive at the explicit expressions
\begin{equation}
\label{eq:stress_update_predictor}
\begin{aligned}
\dev\bfsigma^{t,\text{trial}} &=
2\bar{G} \dev\left(\bfepsilon^t-\boldsymbol{\alpha}_{\RomanNumeralCaps{1}}^{t-1}\right)
- \sum_{j = 1}^{n_{\text{MW}}} 2\frac{G_j}{1+\frac{\Delta t}{g_j}} \dev\boldsymbol{\alpha}_j^{t-1}, \\
\vol\bfsigma^{t,\text{trial}} &=
3\bar{K} \vol\left(\bfepsilon^t-\boldsymbol{\alpha}_{\RomanNumeralCaps{1}}^{t-1}\right)
- \sum_{j = 1}^{n_{\text{MW}}} 3\frac{K_j}{1+\frac{\Delta t}{k_j}} \vol\boldsymbol{\alpha}_j^{t-1},
\end{aligned}
\end{equation}
where we defined
\begin{equation}
\begin{aligned}
\bar{G} &=
G + \sum_{j = 1}^{n_{\text{MW}}} G_j \left( 1 - \frac{\frac{\Delta t}{g_j}}{1+\frac{\Delta t}{g_j}}\right), \\
\bar{K} &=
K + \sum_{j = 1}^{n_{\text{MW}}} K_j \left( 1 - \frac{\frac{\Delta t}{k_j}}{1+\frac{\Delta t}{k_j}}\right),
\end{aligned}
\end{equation}
We can determine whether the viscoelastic predictor is admissible by computing $f^{t,\text{trial}}$ dependent on $\bfsigma^{t,\text{trial}}$. If $f^{t,\text{trial}}>0$, the viscoelastic predictor is not admissible and the viscoplastic corrector step needs to be applied.

\subsection{Viscoplastic corrector step}
\label{sec:viscoplastic_corrector}
The computation of the viscoplastic internal variables in the viscoplastic corrector step can be greatly simplified by taking advantage of some characteristics of the problem. To this end, we define a scalar internal variable $\gamma = \sqrt{\frac{3}{2}} \alpha_{\RomanNumeralCaps{2}}$. After discretization in time and noting that $f^\text{{t,trial}}>0$, the evolution laws in \eqref{eq:evolution_VP} can then be written as
\begin{equation}
\label{eq:evolution_VP_reduced}
\Delta{\boldsymbol{\alpha}}_{\RomanNumeralCaps{1}} = \Delta{\boldsymbol{\alpha}}_{\RomanNumeralCaps{3}}
=\Delta\gamma \frac{\dev\left(\bfsigma^t -H^{kin}\boldsymbol{\alpha}_{\RomanNumeralCaps{3}}^t\right)}{\left\|\dev\left(\bfsigma^t -H^{kin}\boldsymbol{\alpha}_{\RomanNumeralCaps{3}}^t\right)\right\|},
\qquad
\Delta{{\alpha}}_{\RomanNumeralCaps{2}}
=\sqrt{\frac{2}{3}} \Delta\gamma.
\end{equation}
It will be shown that the internal variable update $\Delta\gamma$ can be computed explicitly dependent on the variables at the previous time step, and afterwards, the  update of the other viscoplastic internal variables (i.e., $\Delta\boldsymbol{\alpha}_{\RomanNumeralCaps{1}}=\Delta\boldsymbol{\alpha}_{\RomanNumeralCaps{3}}$ and $\Delta\alpha_{\RomanNumeralCaps{2}}$) can be computed dependent on $\Delta\gamma$. The explicit expression for $\Delta\gamma$ and the relation between $\Delta\gamma$ and the other internal variable updates are derived in the following. The availability of these relations leads to a computationally efficient implementation of the viscoplastic corrector step.

With reference to \eqref{eq:stress_update_predictor}, the deviatoric and volumetric parts of the stress at the current time step are
\begin{equation}
\label{eq:stress_update_corrector}
\begin{aligned}
\dev\bfsigma^t
&=2\bar{G} \dev\left(\bfepsilon^t - \left(\boldsymbol{\alpha}_{\RomanNumeralCaps{1}}^{t,\text{trial}} + \Delta\boldsymbol{\alpha}_{\RomanNumeralCaps{1}}\right)\right)
- \sum_{j = 1}^{n_{\text{MW}}} 2\frac{G_j}{1+\frac{\Delta t}{g_j}} \dev\boldsymbol{\alpha}_j^{t-1}, \\
\vol\bfsigma^t
&=3\bar{K} \vol\left(\bfepsilon^t - \left(\boldsymbol{\alpha}_{\RomanNumeralCaps{1}}^{t,\text{trial}} + \Delta\boldsymbol{\alpha}_{\RomanNumeralCaps{1}}\right)\right)
- \sum_{j = 1}^{n_{\text{MW}}} 3\frac{K_j}{1+\frac{\Delta t}{k_j}} \dev\boldsymbol{\alpha}_j^{t-1},
\end{aligned}
\end{equation}
where we made use of $\boldsymbol{\alpha}_{\RomanNumeralCaps{1}}^{t}=\boldsymbol{\alpha}_{\RomanNumeralCaps{1}}^{t,\text{trial}} + \Delta\boldsymbol{\alpha}_{\RomanNumeralCaps{1}}$. Once again referring to \eqref{eq:stress_update_predictor}, we write \eqref{eq:stress_update_corrector} as
\begin{equation}
\label{eq:stress_update_corrector2}
\begin{aligned}
\dev\bfsigma^t
&=\dev\bfsigma^{t,\text{trial}} - 2\bar{G} \dev\Delta\boldsymbol{\alpha}_{\RomanNumeralCaps{1}}, \\
\vol\bfsigma^t
&=\vol\bfsigma^{t,\text{trial}} - 3\bar{K} \vol\Delta\boldsymbol{\alpha}_{\RomanNumeralCaps{1}},
\end{aligned}
\end{equation}
Knowing that the volumetric part of $\Delta\boldsymbol{\alpha}_{\RomanNumeralCaps{1}}$ vanishes (see \eqref{eq:evolution_VP_reduced}), we observe that $\vol\bfsigma^t=\vol\bfsigma^{t,\text{trial}}$. Therefore, we can focus on the stress deviator in the following. Subtracting the term $\dev \left(H^{kin}\boldsymbol{\alpha}_{\RomanNumeralCaps{3}}^t\right)$ from both sides of the first equation in \eqref{eq:stress_update_corrector2} and using $\Delta\boldsymbol{\alpha}_{\RomanNumeralCaps{1}}=\Delta\boldsymbol{\alpha}_{\RomanNumeralCaps{3}}$ gives
\begin{equation}
\dev\left(\bfsigma^t -H^{kin}\boldsymbol{\alpha}_{\RomanNumeralCaps{3}}^t\right)
= \dev\left(\bfsigma^{t,\text{trial}} -H^{kin}\boldsymbol{\alpha}_{\RomanNumeralCaps{3}}^{t,\text{trial}}\right)
- (2\bar{G} + H^{kin})\dev\Delta\boldsymbol{\alpha}_{\RomanNumeralCaps{1}}.
\end{equation}
After inserting the time-discretized evolution law for $\Delta\boldsymbol{\alpha}_{\RomanNumeralCaps{1}}$ (see \eqref{eq:evolution_VP_reduced}), we obtain
\begin{equation}
\label{eq:stress_update_deviatoric}
\dev\left(\bfsigma^t -H^{kin}\boldsymbol{\alpha}_{\RomanNumeralCaps{3}}^t\right)
= \dev\left(\bfsigma^{t,\text{trial}} -H^{kin}\boldsymbol{\alpha}_{\RomanNumeralCaps{3}}^{t,\text{trial}}\right)
- (2\bar{G} + H^{kin}) \Delta\gamma \frac{\dev\left(\bfsigma^t -H^{kin}\boldsymbol{\alpha}_{\RomanNumeralCaps{3}}^t\right)}{\left\|\dev\left(\bfsigma^t -H^{kin}\boldsymbol{\alpha}_{\RomanNumeralCaps{3}}^t\right)\right\|},
\end{equation}
which can be rearranged to yield
\begin{equation}
\left( 1 + \frac{(2\bar{G} + H^{kin}) \Delta\gamma}{\left\|\dev\left(\bfsigma^t -H^{kin}\boldsymbol{\alpha}_{\RomanNumeralCaps{3}}^t\right)\right\|} \right)
\dev\left(\bfsigma^t -H^{kin}\boldsymbol{\alpha}_{\RomanNumeralCaps{3}}^t\right)
= \dev\left(\bfsigma^{t,\text{trial}} -H^{kin}\boldsymbol{\alpha}_{\RomanNumeralCaps{3}}^{t,\text{trial}}\right).
\end{equation}
This implies that the tensors $\dev\left(\bfsigma^t -H^{kin}\boldsymbol{\alpha}_{\RomanNumeralCaps{3}}^t\right)$ and $\dev\left(\bfsigma^{t,\text{trial}} -H^{kin}\boldsymbol{\alpha}_{\RomanNumeralCaps{3}}^{t,\text{trial}}\right)$ only differ by a scalar factor and hence are collinear, i.e.
\begin{equation}
\label{eq:director}
\frac{\dev\left(\bfsigma^t -H^{kin}\boldsymbol{\alpha}_{\RomanNumeralCaps{3}}^t\right)}{\left\|\dev\left(\bfsigma^t -H^{kin}\boldsymbol{\alpha}_{\RomanNumeralCaps{3}}^t\right)\right\|} =
\frac{\dev\left(\bfsigma^{t,\text{trial}} -H^{kin}\boldsymbol{\alpha}_{\RomanNumeralCaps{3}}^{t,\text{trial}}\right)}{\left\|\dev\left(\bfsigma^{t,\text{trial}} -H^{kin}\boldsymbol{\alpha}_{\RomanNumeralCaps{3}}^{t,\text{trial}}\right)\right\|}.
\end{equation}
The importance of this result is best understood when revisiting \eqref{eq:evolution_VP_reduced}. With this new result, the update of the internal variables $\Delta\boldsymbol{\alpha}_{\RomanNumeralCaps{1}}$ and $\Delta\boldsymbol{\alpha}_{\RomanNumeralCaps{3}}$ can be written as a function of $\Delta\gamma$ and of known variables from the previous time step. Hence, it only remains to compute $\Delta\gamma$, as derived in the following.

Inserting \eqref{eq:director} in \eqref{eq:stress_update_deviatoric}, we obtain
\begin{equation}
\dev\left(\bfsigma^t -H^{kin}\boldsymbol{\alpha}_{\RomanNumeralCaps{3}}^t\right)
= \left( 1 - \frac{(2\bar{G} + H^{kin}) \Delta\gamma}{\left\|\dev\left(\bfsigma^{t,\text{trial}} -H^{kin}\boldsymbol{\alpha}_{\RomanNumeralCaps{3}}^{t,\text{trial}}\right)\right\|} \right)
\dev\left(\bfsigma^{t,\text{trial}} -H^{kin}\boldsymbol{\alpha}_{\RomanNumeralCaps{3}}^{t,\text{trial}}\right).
\end{equation}
Using this result, we can write
\begin{equation}
\begin{aligned}
\Delta\gamma &=
\sqrt{\frac{3}{2}} \frac{\Delta t}{\eta_p} f^t, \\
&=
\sqrt{\frac{3}{2}} \frac{\Delta t}{\eta_p} \left( \sqrt{\frac{3}{2}}\left\|\dev\left(\bfsigma^t -H^{kin}\boldsymbol{\alpha}_{\RomanNumeralCaps{3}}^t\right)\right\| - \sigma_{y} -H^{iso}\alpha_{\RomanNumeralCaps{2}}^{t,\text{trial}} -H^{iso}\sqrt{\frac{2}{3}}\Delta\gamma \right), \\
&=
\sqrt{\frac{3}{2}} \frac{\Delta t}{\eta_p} \left( \sqrt{\frac{3}{2}}\left\|\dev\left(\bfsigma^{t,\text{trial}} -H^{kin}\boldsymbol{\alpha}_{\RomanNumeralCaps{3}}^{t,\text{trial}}\right)\right\| - \sqrt{\frac{3}{2}}(2\bar{G} + H^{kin}) \Delta\gamma - \sigma_{y} -H^{iso}\alpha_{\RomanNumeralCaps{2}}^{t,\text{trial}} -H^{iso}\sqrt{\frac{2}{3}}\Delta\gamma \right), \\
&=
\sqrt{\frac{3}{2}} \frac{\Delta t}{\eta_p} \left( f^{t,\text{trial}} - \sqrt{\frac{3}{2}}(2\bar{G} + H^{kin}) \Delta\gamma -H^{iso}\sqrt{\frac{2}{3}}\Delta\gamma \right),
\end{aligned}
\end{equation}
which is solved for $\Delta\gamma$ to obtain
\begin{equation}
\Delta\gamma = \frac{f^{t,\text{trial}}}{\sqrt{\frac{2}{3}} \left(\frac{\eta_p}{\Delta t} + H^{iso}\right) + \sqrt{\frac{3}{2}}(2\bar{G} + H^{kin})}.
\end{equation}
After computing $\Delta\gamma$, the other internal variables (see \eqref{eq:evolution_VP_reduced} and \eqref{eq:director}) and the stress (see \eqref{eq:stress_update_corrector}) can be computed.

\section{Consistent tangent}
\label{sec:consisten_tangent}
Implementing the material model library in forward finite element simulations requires the computation of the consistent tangent
\begin{equation}
\mathbb{C}^t = \frac{\partial \bfsigma^t}{\partial\bfepsilon^t}.
\end{equation}
We consider two different cases. If the viscoelastic predictor in \ref{sec:viscoelastic_predictor} is admissible, it is
\begin{equation}
\mathbb{C}^t = \mathbb{C}^{t,\text{trial}} = \frac{\partial \bfsigma^{t,\text{trial}}}{\partial\bfepsilon^t} = 2\bar{G} \mathbb{I}^{\text{dev}} + 3\bar{K} \mathbb{I}^{\text{vol}}.
\end{equation}
Otherwise, we obtain
\begin{equation}
\begin{aligned}
\mathbb{C}^t
&=\left( 1 - \frac{2\bar{G}\Delta\gamma}{\left\|\dev\left(\bfsigma^{t,\text{trial}} -H^{kin}\boldsymbol{\alpha}_{\RomanNumeralCaps{3}}^{t,\text{trial}}\right)\right\|} \right) 2\bar{G} \mathbb{I}^{\text{dev}}
+ 3\bar{K} \mathbb{I}^{\text{vol}} \\
&- \frac{2\bar{G}\Delta\gamma\left( \sigma_{y} + H^{iso}\alpha_{\RomanNumeralCaps{2}}^{t,\text{trial}} \right)}{f^{t,\text{trial}}\left\|\dev\left(\bfsigma^{t,\text{trial}} -H^{kin}\boldsymbol{\alpha}_{\RomanNumeralCaps{3}}^{t,\text{trial}}\right)\right\|^3}
\mathbb{C}^{t,\text{trial}}
\colon \left( \dev\left(\bfsigma^{t,\text{trial}} -H^{kin}\boldsymbol{\alpha}_{\RomanNumeralCaps{3}}^{t,\text{trial}}\right) \otimes \dev\left(\bfsigma^{t,\text{trial}} -H^{kin}\boldsymbol{\alpha}_{\RomanNumeralCaps{3}}^{t,\text{trial}}\right) \right).
\end{aligned}
\end{equation}
Note that the consistent tangent is required for the solution of the forward finite element problem but not for the inverse problem, i.e., not for EUCLID.

\section{Numerical settings}
\label{sec:settings}
\tablename\ref{tab:settings} provides a list of parameters and hyperparameters used during the data generation via finite element simulations and during the inverse discovery process via EUCLID.

\begin{table}
	\caption{(Hyper-)parameters for the finite element simulations and EUCLID.}\label{tab:settings}
	\centering
	\begin{tabular}{lccc}
	    \hline
		Parameter & Notation & Value & Unit \\ \hline
		Number of nodes in mesh & $n_n$ & $2,179$ & - \\
		Number of measured reaction force components & $n_\idxREAC$ & $2$ & - \\
		Number of time steps & $n_t$ & $800$ & - \\
		Number of loading phases & $n_{\idxPHASE}$ & $20$ & - \\
		Applied loading per period & $\delta$ & $\pm 0.5$ & mm \\
		Time per loading period & $t^{\text{load}}_{\idxPHASE}$ & $\{ 10^{ ( -2 + \frac{4}{n_{\idxPHASE}-1} i ) } : i=0,\dots,(n_{\idxPHASE}-1)\}$ & s \\
		Time per relaxation period & $t^{\text{relax}}_{\idxPHASE}$ & $10^{2}$ & s \\
		Displacement noise standard deviation & $\sigma$ & $\left\{0,10^{-4},3 \cdot 10^{-4},5 \cdot 10^{-4}\right\}$ & mm\\
		Moving-window length for denoising & - & $10$ & -\\
		Coefficient for reaction force balance & $\lambda_r$ & 100 & - \\
		Number of random initial guesses & $n_g$ & $24$ & -\\
		Exponent for $L_p$ regularization & $p$ & $1$ & - \\
		Number of choices of $\lambda_p$ & $n_\lambda$ & $24$ & -\\
		Coefficient for $L_p$ regularization & $\lambda_p$ & $\{10^{-4} \cdot 2^i : i=0,\dots,23\}$ & $\text{kN}^2$ \\
		Lower bound for $\bftheta$ & $\theta^{\text{min}}_i$ & $\{ 0, \theta^{\text{lb}} \}$ & various\\
		Lower bound for $\bftheta$ (reciprocals) & $\theta^{\text{lb}}$ & $10^{-6}$ & various\\
		Threshold for $C$ & $C^{\text{th}}$ & $\max \{ 10^{-5} , 1.1 \ C^{\text{min}} \}$ & $\text{kN}^2$\\
		Threshold for $\bftheta$ & $\theta^{\text{th}}$ & $10^{-4}$ & various\\
		
		\hline
	\end{tabular}
\end{table}


\section*{Code and data availability}

Codes and data are publicly available at \url{https://euclid-code.github.io/}.
\bibliographystyle{elsarticle-harv}
\bibliography{EUCLID-GSM}

\begin{thebibliography}{81}
\expandafter\ifx\csname natexlab\endcsname\relax\def\natexlab#1{#1}\fi
\providecommand{\url}[1]{\texttt{#1}}
\providecommand{\href}[2]{#2}
\providecommand{\path}[1]{#1}
\providecommand{\DOIprefix}{doi:}
\providecommand{\ArXivprefix}{arXiv:}
\providecommand{\URLprefix}{URL: }
\providecommand{\Pubmedprefix}{pmid:}
\providecommand{\doi}[1]{\href{http://dx.doi.org/#1}{\path{#1}}}
\providecommand{\Pubmed}[1]{\href{pmid:#1}{\path{#1}}}
\providecommand{\bibinfo}[2]{#2}
\ifx\xfnm\relax \def\xfnm[#1]{\unskip,\space#1}\fi
\bibitem[{Amores et~al.(2022)Amores, Montáns, Cueto and
  Chinesta}]{amores_crossing_2022}
\bibinfo{author}{Amores, V.J.}, \bibinfo{author}{Montáns, F.J.},
  \bibinfo{author}{Cueto, E.}, \bibinfo{author}{Chinesta, F.},
  \bibinfo{year}{2022}.
\newblock \bibinfo{title}{Crossing {Scales}: {Data}-{Driven} {Determination} of
  the {Micro}-scale {Behavior} of {Polymers} {From} {Non}-homogeneous {Tests}
  at the {Continuum}-{Scale}}.
\newblock \bibinfo{journal}{Frontiers in Materials} \bibinfo{volume}{9},
  \bibinfo{pages}{879614}.
\newblock \URLprefix
  \url{https://www.frontiersin.org/articles/10.3389/fmats.2022.879614/full},
  \DOIprefix\doi{10.3389/fmats.2022.879614}.
\bibitem[{Anton and Wessels(2022)}]{anton_identification_2022}
\bibinfo{author}{Anton, D.}, \bibinfo{author}{Wessels, H.},
  \bibinfo{year}{2022}.
\newblock \bibinfo{title}{Identification of {Material} {Parameters} from
  {Full}-{Field} {Displacement} {Data} {Using} {Physics}-{Informed} {Neural}
  {Networks}} .
\bibitem[{As'ad et~al.(2022)As'ad, Avery and
  Farhat}]{asad_mechanics-informed_2022}
\bibinfo{author}{As'ad, F.}, \bibinfo{author}{Avery, P.},
  \bibinfo{author}{Farhat, C.}, \bibinfo{year}{2022}.
\newblock \bibinfo{title}{A {Mechanics}-{Informed} {Artificial} {Neural}
  {Network} {Approach} in {Data}-{Driven} {Constitutive} {Modeling}}, in:
  \bibinfo{booktitle}{{AIAA} {SCITECH} 2022 {Forum}},
  \bibinfo{publisher}{American Institute of Aeronautics and Astronautics},
  \bibinfo{address}{San Diego, CA \& Virtual}.
\newblock \URLprefix \url{https://arc.aiaa.org/doi/10.2514/6.2022-0100},
  \DOIprefix\doi{10.2514/6.2022-0100}.
\bibitem[{Avril et~al.(2008)Avril, Bonnet, Bretelle, Grédiac, Hild, Ienny,
  Latourte, Lemosse, Pagano, Pagnacco and Pierron}]{avril_overview_2008}
\bibinfo{author}{Avril, S.}, \bibinfo{author}{Bonnet, M.},
  \bibinfo{author}{Bretelle, A.S.}, \bibinfo{author}{Grédiac, M.},
  \bibinfo{author}{Hild, F.}, \bibinfo{author}{Ienny, P.},
  \bibinfo{author}{Latourte, F.}, \bibinfo{author}{Lemosse, D.},
  \bibinfo{author}{Pagano, S.}, \bibinfo{author}{Pagnacco, E.},
  \bibinfo{author}{Pierron, F.}, \bibinfo{year}{2008}.
\newblock \bibinfo{title}{Overview of {Identification} {Methods} of
  {Mechanical} {Parameters} {Based} on {Full}-field {Measurements}}.
\newblock \bibinfo{journal}{Experimental Mechanics} \bibinfo{volume}{48},
  \bibinfo{pages}{381--402}.
\newblock \URLprefix \url{http://link.springer.com/10.1007/s11340-008-9148-y},
  \DOIprefix\doi{10.1007/s11340-008-9148-y}.
\bibitem[{Blühdorn et~al.(2022)Blühdorn, Gauger and
  Kabel}]{bluhdorn_automat_2022}
\bibinfo{author}{Blühdorn, J.}, \bibinfo{author}{Gauger, N.R.},
  \bibinfo{author}{Kabel, M.}, \bibinfo{year}{2022}.
\newblock \bibinfo{title}{{AutoMat}: automatic differentiation for generalized
  standard materials on {GPUs}}.
\newblock \bibinfo{journal}{Computational Mechanics} \bibinfo{volume}{69},
  \bibinfo{pages}{589--613}.
\newblock \URLprefix
  \url{https://link.springer.com/10.1007/s00466-021-02105-2},
  \DOIprefix\doi{10.1007/s00466-021-02105-2}.
\bibitem[{Brunton et~al.(2016)Brunton, Proctor and
  Kutz}]{brunton_discovering_2016}
\bibinfo{author}{Brunton, S.L.}, \bibinfo{author}{Proctor, J.L.},
  \bibinfo{author}{Kutz, J.N.}, \bibinfo{year}{2016}.
\newblock \bibinfo{title}{Discovering governing equations from data by sparse
  identification of nonlinear dynamical systems}.
\newblock \bibinfo{journal}{Proceedings of the National Academy of Sciences}
  \bibinfo{volume}{113}, \bibinfo{pages}{3932--3937}.
\newblock \URLprefix
  \url{http://www.pnas.org/lookup/doi/10.1073/pnas.1517384113},
  \DOIprefix\doi{10.1073/pnas.1517384113}.
\bibitem[{Bröcker and Matzenmiller(2012)}]{brocker_thermoviscoplasticity_2012}
\bibinfo{author}{Bröcker, C.}, \bibinfo{author}{Matzenmiller, A.},
  \bibinfo{year}{2012}.
\newblock \bibinfo{title}{Thermoviscoplasticity deduced from enhanced
  rheological models}.
\newblock \bibinfo{journal}{PAMM} \bibinfo{volume}{12},
  \bibinfo{pages}{327--328}.
\newblock \URLprefix
  \url{https://onlinelibrary.wiley.com/doi/10.1002/pamm.201210152},
  \DOIprefix\doi{10.1002/pamm.201210152}.
\bibitem[{Dalémat et~al.(2019)Dalémat, Coret, Leygue and
  Verron}]{dalemat_measuring_2019}
\bibinfo{author}{Dalémat, M.}, \bibinfo{author}{Coret, M.},
  \bibinfo{author}{Leygue, A.}, \bibinfo{author}{Verron, E.},
  \bibinfo{year}{2019}.
\newblock \bibinfo{title}{Measuring stress field without constitutive
  equation}.
\newblock \bibinfo{journal}{Mechanics of Materials} \bibinfo{volume}{136},
  \bibinfo{pages}{103087}.
\newblock \DOIprefix\doi{10.1016/j.mechmat.2019.103087}.
\bibitem[{Doi(2011)}]{doi_onsagers_2011}
\bibinfo{author}{Doi, M.}, \bibinfo{year}{2011}.
\newblock \bibinfo{title}{Onsager’s variational principle in soft matter}.
\newblock \bibinfo{journal}{Journal of Physics: Condensed Matter}
  \bibinfo{volume}{23}, \bibinfo{pages}{284118}.
\newblock \URLprefix
  \url{https://iopscience.iop.org/article/10.1088/0953-8984/23/28/284118},
  \DOIprefix\doi{10.1088/0953-8984/23/28/284118}.
\bibitem[{Flaschel et~al.(2021)Flaschel, Kumar and
  De~Lorenzis}]{flaschel_unsupervised_2021}
\bibinfo{author}{Flaschel, M.}, \bibinfo{author}{Kumar, S.},
  \bibinfo{author}{De~Lorenzis, L.}, \bibinfo{year}{2021}.
\newblock \bibinfo{title}{Unsupervised discovery of interpretable hyperelastic
  constitutive laws}.
\newblock \bibinfo{journal}{Computer Methods in Applied Mechanics and
  Engineering} \bibinfo{volume}{381}, \bibinfo{pages}{113852}.
\newblock \DOIprefix\doi{10.1016/j.cma.2021.113852}.
\bibitem[{Flaschel et~al.(2022)Flaschel, Kumar and
  De~Lorenzis}]{flaschel_discovering_2022}
\bibinfo{author}{Flaschel, M.}, \bibinfo{author}{Kumar, S.},
  \bibinfo{author}{De~Lorenzis, L.}, \bibinfo{year}{2022}.
\newblock \bibinfo{title}{Discovering plasticity models without stress data}.
\newblock \bibinfo{journal}{npj Computational Materials} \bibinfo{volume}{8},
  \bibinfo{pages}{91}.
\newblock \URLprefix \url{https://www.nature.com/articles/s41524-022-00752-4},
  \DOIprefix\doi{10.1038/s41524-022-00752-4}.
\bibitem[{Frank and Friedman(1993)}]{frank_statistical_1993}
\bibinfo{author}{Frank, l.E.}, \bibinfo{author}{Friedman, J.H.},
  \bibinfo{year}{1993}.
\newblock \bibinfo{title}{A {Statistical} {View} of {Some} {Chemometrics}
  {Regression} {Tools}}.
\newblock \bibinfo{journal}{Technometrics} \bibinfo{volume}{35},
  \bibinfo{pages}{109--135}.
\newblock \URLprefix
  \url{http://www.tandfonline.com/doi/abs/10.1080/00401706.1993.10485033},
  \DOIprefix\doi{10.1080/00401706.1993.10485033}.
\bibitem[{Frankel et~al.(2020)Frankel, Jones and Swiler}]{frankel_tensor_2020}
\bibinfo{author}{Frankel, A.L.}, \bibinfo{author}{Jones, R.E.},
  \bibinfo{author}{Swiler, L.P.}, \bibinfo{year}{2020}.
\newblock \bibinfo{title}{{TENSOR} {BASIS} {GAUSSIAN} {PROCESS} {MODELS} {OF}
  {HYPERELASTIC} {MATERIALS}}.
\newblock \bibinfo{journal}{Journal of Machine Learning for Modeling and
  Computing} \bibinfo{volume}{1}, \bibinfo{pages}{1--17}.
\newblock \URLprefix
  \url{http://www.dl.begellhouse.com/journals/558048804a15188a,583c4e56625ba94e,651a2e6b0260f708.html},
  \DOIprefix\doi{10.1615/JMachLearnModelComput.2020033325}.
\bibitem[{Fuhg et~al.(2021)Fuhg, Boehm, Bouklas, Fau, Wriggers and
  Marino}]{fuhg_model-data-driven_2021}
\bibinfo{author}{Fuhg, J.N.}, \bibinfo{author}{Boehm, C.},
  \bibinfo{author}{Bouklas, N.}, \bibinfo{author}{Fau, A.},
  \bibinfo{author}{Wriggers, P.}, \bibinfo{author}{Marino, M.},
  \bibinfo{year}{2021}.
\newblock \bibinfo{title}{Model-data-driven constitutive responses: application
  to a multiscale computational framework}.
\newblock \bibinfo{journal}{arXiv:2104.02650 [cs, math, stat]} \URLprefix
  \url{http://arxiv.org/abs/2104.02650}. \bibinfo{note}{arXiv: 2104.02650}.
\bibitem[{Fuhg et~al.(2022)Fuhg, Fau, Bouklas and
  Marino}]{fuhg_elasto-plasticity_2022}
\bibinfo{author}{Fuhg, J.N.}, \bibinfo{author}{Fau, A.},
  \bibinfo{author}{Bouklas, N.}, \bibinfo{author}{Marino, M.},
  \bibinfo{year}{2022}.
\newblock \bibinfo{title}{Elasto-plasticity with convex model-data-driven yield
  functions} , \bibinfo{pages}{27}.
\bibitem[{Ghaboussi et~al.(1991)Ghaboussi, Garrett and
  Wu}]{ghaboussi_knowledgebased_1991}
\bibinfo{author}{Ghaboussi, J.}, \bibinfo{author}{Garrett, J.H.},
  \bibinfo{author}{Wu, X.}, \bibinfo{year}{1991}.
\newblock \bibinfo{title}{Knowledge‐{Based} {Modeling} of {Material}
  {Behavior} with {Neural} {Networks}}.
\newblock \bibinfo{journal}{Journal of Engineering Mechanics}
  \bibinfo{volume}{117}, \bibinfo{pages}{132--153}.
\newblock \URLprefix
  \url{http://ascelibrary.org/doi/10.1061/%28ASCE%290733-9399%281991%29117%3A1%28132%29},
  \DOIprefix\doi{10.1061/(ASCE)0733-9399(1991)117:1(132)}.
\bibitem[{Giunta and Angela~Pisano(2006)}]{giunta_onedimensional_2006}
\bibinfo{author}{Giunta, M.}, \bibinfo{author}{Angela~Pisano, A.},
  \bibinfo{year}{2006}.
\newblock \bibinfo{title}{One‐{Dimensional} {Visco}‐{Elastoplastic}
  {Constitutive} {Model} for {Asphalt} {Concrete}}.
\newblock \bibinfo{journal}{Multidiscipline Modeling in Materials and
  Structures} \bibinfo{volume}{2}, \bibinfo{pages}{247--264}.
\newblock \URLprefix
  \url{https://www.emerald.com/insight/content/doi/10.1163/157361106776240761/full/html},
  \DOIprefix\doi{10.1163/157361106776240761}.
\bibitem[{González et~al.(2019a)González, Chinesta and
  Cueto}]{gonzalez_learning_2019}
\bibinfo{author}{González, D.}, \bibinfo{author}{Chinesta, F.},
  \bibinfo{author}{Cueto, E.}, \bibinfo{year}{2019}a.
\newblock \bibinfo{title}{Learning {Corrections} for {Hyperelastic} {Models}
  {From} {Data}}.
\newblock \bibinfo{journal}{Frontiers in Materials} \bibinfo{volume}{6},
  \bibinfo{pages}{14}.
\newblock \URLprefix
  \url{https://www.frontiersin.org/article/10.3389/fmats.2019.00014/full},
  \DOIprefix\doi{10.3389/fmats.2019.00014}.
\bibitem[{González et~al.(2019b)González, Chinesta and
  Cueto}]{gonzalez_thermodynamically_2019}
\bibinfo{author}{González, D.}, \bibinfo{author}{Chinesta, F.},
  \bibinfo{author}{Cueto, E.}, \bibinfo{year}{2019}b.
\newblock \bibinfo{title}{Thermodynamically consistent data-driven
  computational mechanics}.
\newblock \bibinfo{journal}{Continuum Mechanics and Thermodynamics}
  \bibinfo{volume}{31}, \bibinfo{pages}{239--253}.
\newblock \URLprefix \url{http://link.springer.com/10.1007/s00161-018-0677-z},
  \DOIprefix\doi{10.1007/s00161-018-0677-z}.
\bibitem[{Grmela and Öttinger(1997)}]{grmela_dynamics_1997}
\bibinfo{author}{Grmela, M.}, \bibinfo{author}{Öttinger, H.C.},
  \bibinfo{year}{1997}.
\newblock \bibinfo{title}{Dynamics and thermodynamics of complex fluids. {I}.
  {Development} of a general formalism}.
\newblock \bibinfo{journal}{Physical Review E} \bibinfo{volume}{56},
  \bibinfo{pages}{6620--6632}.
\newblock \URLprefix \url{https://link.aps.org/doi/10.1103/PhysRevE.56.6620},
  \DOIprefix\doi{10.1103/PhysRevE.56.6620}.
\bibitem[{Grédiac(1989)}]{grediac_principle_1989}
\bibinfo{author}{Grédiac, M.}, \bibinfo{year}{1989}.
\newblock \bibinfo{title}{Principle of virtual work and identification}.
\newblock \bibinfo{journal}{Comptes Rendus de L Academie des Sciences Serie Ii}
  , \bibinfo{pages}{1--5}.
\bibitem[{Halphen and Nguyen(1975)}]{halphen_sur_1975}
\bibinfo{author}{Halphen, B.}, \bibinfo{author}{Nguyen, Q.S.},
  \bibinfo{year}{1975}.
\newblock \bibinfo{title}{Sur les matériaux standard généralisés} ,
  \bibinfo{pages}{26}.
\bibitem[{Hartmann(2001)}]{hartmann_numerical_2001}
\bibinfo{author}{Hartmann, S.}, \bibinfo{year}{2001}.
\newblock \bibinfo{title}{Numerical studies on the identification of the
  material parameters of {Rivlin}'s hyperelasticity using tension-torsion
  tests}.
\newblock \bibinfo{journal}{Acta Mechanica} \bibinfo{volume}{148},
  \bibinfo{pages}{129--155}.
\newblock \URLprefix \url{http://link.springer.com/10.1007/BF01183674},
  \DOIprefix\doi{10.1007/BF01183674}.
\bibitem[{Hernández et~al.(2021)Hernández, Badías, González, Chinesta and
  Cueto}]{hernandez_structure-preserving_2021}
\bibinfo{author}{Hernández, Q.}, \bibinfo{author}{Badías, A.},
  \bibinfo{author}{González, D.}, \bibinfo{author}{Chinesta, F.},
  \bibinfo{author}{Cueto, E.}, \bibinfo{year}{2021}.
\newblock \bibinfo{title}{Structure-preserving neural networks}.
\newblock \bibinfo{journal}{Journal of Computational Physics}
  \bibinfo{volume}{426}, \bibinfo{pages}{109950}.
\newblock \URLprefix
  \url{https://linkinghub.elsevier.com/retrieve/pii/S0021999120307245},
  \DOIprefix\doi{10.1016/j.jcp.2020.109950}.
\bibitem[{Hild and Roux(2006)}]{hild_digital_2006}
\bibinfo{author}{Hild, F.}, \bibinfo{author}{Roux, S.}, \bibinfo{year}{2006}.
\newblock \bibinfo{title}{Digital {Image} {Correlation}: from {Displacement}
  {Measurement} to {Identification} of {Elastic} {Properties} - a {Review}}.
\newblock \bibinfo{journal}{Strain} \bibinfo{volume}{42},
  \bibinfo{pages}{69--80}.
\newblock \URLprefix
  \url{http://doi.wiley.com/10.1111/j.1475-1305.2006.00258.x},
  \DOIprefix\doi{10.1111/j.1475-1305.2006.00258.x}.
\bibitem[{Huang et~al.(2020)Huang, Xu, Farhat and Darve}]{huang_learning_2020}
\bibinfo{author}{Huang, D.Z.}, \bibinfo{author}{Xu, K.},
  \bibinfo{author}{Farhat, C.}, \bibinfo{author}{Darve, E.},
  \bibinfo{year}{2020}.
\newblock \bibinfo{title}{Learning constitutive relations from indirect
  observations using deep neural networks}.
\newblock \bibinfo{journal}{Journal of Computational Physics}
  \bibinfo{volume}{416}, \bibinfo{pages}{109491}.
\newblock \URLprefix
  \url{https://linkinghub.elsevier.com/retrieve/pii/S0021999120302655},
  \DOIprefix\doi{10.1016/j.jcp.2020.109491}.
\bibitem[{Huang et~al.(2022)Huang, He, Chem and Reina}]{huang_variational_2022}
\bibinfo{author}{Huang, S.}, \bibinfo{author}{He, Z.}, \bibinfo{author}{Chem,
  B.}, \bibinfo{author}{Reina, C.}, \bibinfo{year}{2022}.
\newblock \bibinfo{title}{Variational {Onsager} {Neural} {Networks} ({VONNs}):
  {A} thermodynamics-based variational learning strategy for non-equilibrium
  {PDEs}}.
\newblock \bibinfo{journal}{Journal of the Mechanics and Physics of Solids}
  \bibinfo{volume}{163}, \bibinfo{pages}{104856}.
\newblock \URLprefix
  \url{https://linkinghub.elsevier.com/retrieve/pii/S0022509622000692},
  \DOIprefix\doi{10.1016/j.jmps.2022.104856}.
\bibitem[{Hütter and Svendsen(2011)}]{hutter_formulation_2011}
\bibinfo{author}{Hütter, M.}, \bibinfo{author}{Svendsen, B.},
  \bibinfo{year}{2011}.
\newblock \bibinfo{title}{On the {Formulation} of {Continuum} {Thermodynamic}
  {Models} for {Solids} as {General} {Equations} for {Non}-equilibrium
  {Reversible}-{Irreversible} {Coupling}} , \bibinfo{pages}{12}.
\bibitem[{Ibañez et~al.(2018)Ibañez, Abisset-Chavanne, Aguado, Gonzalez,
  Cueto and Chinesta}]{ibanez_manifold_2018}
\bibinfo{author}{Ibañez, R.}, \bibinfo{author}{Abisset-Chavanne, E.},
  \bibinfo{author}{Aguado, J.V.}, \bibinfo{author}{Gonzalez, D.},
  \bibinfo{author}{Cueto, E.}, \bibinfo{author}{Chinesta, F.},
  \bibinfo{year}{2018}.
\newblock \bibinfo{title}{A {Manifold} {Learning} {Approach} to {Data}-{Driven}
  {Computational} {Elasticity} and {Inelasticity}}.
\newblock \bibinfo{journal}{Archives of Computational Methods in Engineering}
  \bibinfo{volume}{25}, \bibinfo{pages}{47--57}.
\newblock \URLprefix \url{http://link.springer.com/10.1007/s11831-016-9197-9},
  \DOIprefix\doi{10.1007/s11831-016-9197-9}.
\bibitem[{Ibáñez et~al.(2019)Ibáñez, Abisset-Chavanne, González, Duval,
  Cueto and Chinesta}]{ibanez_hybrid_2019}
\bibinfo{author}{Ibáñez, R.}, \bibinfo{author}{Abisset-Chavanne, E.},
  \bibinfo{author}{González, D.}, \bibinfo{author}{Duval, J.L.},
  \bibinfo{author}{Cueto, E.}, \bibinfo{author}{Chinesta, F.},
  \bibinfo{year}{2019}.
\newblock \bibinfo{title}{Hybrid constitutive modeling: data-driven learning of
  corrections to plasticity models}.
\newblock \bibinfo{journal}{International Journal of Material Forming}
  \bibinfo{volume}{12}, \bibinfo{pages}{717--725}.
\newblock \URLprefix \url{http://link.springer.com/10.1007/s12289-018-1448-x},
  \DOIprefix\doi{10.1007/s12289-018-1448-x}.
\bibitem[{Joshi et~al.(2022)Joshi, Thakolkaran, Zheng, Escande, Flaschel,
  De~Lorenzis and Kumar}]{joshi_bayesian-euclid_2022}
\bibinfo{author}{Joshi, A.}, \bibinfo{author}{Thakolkaran, P.},
  \bibinfo{author}{Zheng, Y.}, \bibinfo{author}{Escande, M.},
  \bibinfo{author}{Flaschel, M.}, \bibinfo{author}{De~Lorenzis, L.},
  \bibinfo{author}{Kumar, S.}, \bibinfo{year}{2022}.
\newblock \bibinfo{title}{Bayesian-{EUCLID}: {Discovering} hyperelastic
  material laws with uncertainties}.
\newblock \bibinfo{journal}{Computer Methods in Applied Mechanics and
  Engineering} \bibinfo{volume}{398}, \bibinfo{pages}{115225}.
\newblock \URLprefix
  \url{https://linkinghub.elsevier.com/retrieve/pii/S0045782522003681},
  \DOIprefix\doi{10.1016/j.cma.2022.115225}.
\bibitem[{Kalina et~al.(2022)Kalina, Linden, Brummund, Metsch and
  Kästner}]{kalina_automated_2022}
\bibinfo{author}{Kalina, K.A.}, \bibinfo{author}{Linden, L.},
  \bibinfo{author}{Brummund, J.}, \bibinfo{author}{Metsch, P.},
  \bibinfo{author}{Kästner, M.}, \bibinfo{year}{2022}.
\newblock \bibinfo{title}{Automated constitutive modeling of isotropic
  hyperelasticity based on artificial neural networks}.
\newblock \bibinfo{journal}{Computational Mechanics} \bibinfo{volume}{69},
  \bibinfo{pages}{213--232}.
\newblock \URLprefix
  \url{https://link.springer.com/10.1007/s00466-021-02090-6},
  \DOIprefix\doi{10.1007/s00466-021-02090-6}.
\bibitem[{Kirchdoerfer and Ortiz(2016)}]{kirchdoerfer_data-driven_2016}
\bibinfo{author}{Kirchdoerfer, T.}, \bibinfo{author}{Ortiz, M.},
  \bibinfo{year}{2016}.
\newblock \bibinfo{title}{Data-driven computational mechanics}.
\newblock \bibinfo{journal}{Computer Methods in Applied Mechanics and
  Engineering} \bibinfo{volume}{304}, \bibinfo{pages}{81--101}.
\newblock \URLprefix
  \url{https://linkinghub.elsevier.com/retrieve/pii/S0045782516300238},
  \DOIprefix\doi{10.1016/j.cma.2016.02.001}.
\bibitem[{Klein et~al.(2022a)Klein, Fernández, Martin, Neff and
  Weeger}]{klein_polyconvex_2022}
\bibinfo{author}{Klein, D.K.}, \bibinfo{author}{Fernández, M.},
  \bibinfo{author}{Martin, R.J.}, \bibinfo{author}{Neff, P.},
  \bibinfo{author}{Weeger, O.}, \bibinfo{year}{2022}a.
\newblock \bibinfo{title}{Polyconvex anisotropic hyperelasticity with neural
  networks}.
\newblock \bibinfo{journal}{Journal of the Mechanics and Physics of Solids}
  \bibinfo{volume}{159}, \bibinfo{pages}{104703}.
\newblock \URLprefix
  \url{https://linkinghub.elsevier.com/retrieve/pii/S0022509621003215},
  \DOIprefix\doi{10.1016/j.jmps.2021.104703}.
\bibitem[{Klein et~al.(2022b)Klein, Ortigosa, Martínez-Frutos and
  Weeger}]{klein_finite_2022}
\bibinfo{author}{Klein, D.K.}, \bibinfo{author}{Ortigosa, R.},
  \bibinfo{author}{Martínez-Frutos, J.}, \bibinfo{author}{Weeger, O.},
  \bibinfo{year}{2022}b.
\newblock \bibinfo{title}{Finite electro-elasticity with physics-augmented
  neural networks}.
\newblock \bibinfo{journal}{Computer Methods in Applied Mechanics and
  Engineering} \bibinfo{volume}{400}, \bibinfo{pages}{115501}.
\newblock \URLprefix
  \url{https://linkinghub.elsevier.com/retrieve/pii/S004578252200514X},
  \DOIprefix\doi{10.1016/j.cma.2022.115501}.
\bibitem[{Korelc(2002)}]{korelc_multi-language_2002}
\bibinfo{author}{Korelc, J.}, \bibinfo{year}{2002}.
\newblock \bibinfo{title}{Multi-language and {Multi}-environment {Generation}
  of {Nonlinear} {Finite} {Element} {Codes}}.
\newblock \bibinfo{journal}{Engineering with Computers} \bibinfo{volume}{18},
  \bibinfo{pages}{312--327}.
\newblock \URLprefix \url{http://link.springer.com/10.1007/s003660200028},
  \DOIprefix\doi{10.1007/s003660200028}.
\bibitem[{Kumar and Lopez-Pamies(2016)}]{kumar_two-potential_2016}
\bibinfo{author}{Kumar, A.}, \bibinfo{author}{Lopez-Pamies, O.},
  \bibinfo{year}{2016}.
\newblock \bibinfo{title}{On the two-potential constitutive modeling of rubber
  viscoelastic materials}.
\newblock \bibinfo{journal}{Comptes Rendus Mécanique} \bibinfo{volume}{344},
  \bibinfo{pages}{102--112}.
\newblock \URLprefix
  \url{https://linkinghub.elsevier.com/retrieve/pii/S1631072115001448},
  \DOIprefix\doi{10.1016/j.crme.2015.11.004}.
\bibitem[{Kumar and Kochmann(2022)}]{aldakheel_what_2022}
\bibinfo{author}{Kumar, S.}, \bibinfo{author}{Kochmann, D.M.},
  \bibinfo{year}{2022}.
\newblock \bibinfo{title}{What {Machine} {Learning} {Can} {Do} for
  {Computational} {Solid} {Mechanics}}, in: \bibinfo{editor}{Aldakheel, F.},
  \bibinfo{editor}{Hudobivnik, B.}, \bibinfo{editor}{Soleimani, M.},
  \bibinfo{editor}{Wessels, H.}, \bibinfo{editor}{Weißenfels, C.},
  \bibinfo{editor}{Marino, M.} (Eds.), \bibinfo{booktitle}{Current {Trends} and
  {Open} {Problems} in {Computational} {Mechanics}}.
  \bibinfo{publisher}{Springer International Publishing},
  \bibinfo{address}{Cham}, pp. \bibinfo{pages}{275--285}.
\newblock \URLprefix
  \url{https://link.springer.com/10.1007/978-3-030-87312-7_27},
  \DOIprefix\doi{10.1007/978-3-030-87312-7_27}.
\bibitem[{Lemaitre and Chaboche(1994)}]{lemaitre_mechanics_1994}
\bibinfo{author}{Lemaitre, J.}, \bibinfo{author}{Chaboche, J.L.},
  \bibinfo{year}{1994}.
\newblock \bibinfo{title}{Mechanics of solid materials}.
\newblock \bibinfo{publisher}{Cambridge university press}.
\bibitem[{Leygue et~al.(2018)Leygue, Coret, Réthoré, Stainier and
  Verron}]{leygue_data-based_2018}
\bibinfo{author}{Leygue, A.}, \bibinfo{author}{Coret, M.},
  \bibinfo{author}{Réthoré, J.}, \bibinfo{author}{Stainier, L.},
  \bibinfo{author}{Verron, E.}, \bibinfo{year}{2018}.
\newblock \bibinfo{title}{Data-based derivation of material response}.
\newblock \bibinfo{journal}{Computer Methods in Applied Mechanics and
  Engineering} \bibinfo{volume}{331}, \bibinfo{pages}{184--196}.
\newblock \URLprefix
  \url{https://linkinghub.elsevier.com/retrieve/pii/S0045782517307156},
  \DOIprefix\doi{10.1016/j.cma.2017.11.013}.
\bibitem[{Liu et~al.(2020)Liu, Tao, Du, Yu and Xu}]{liu_learning_2020}
\bibinfo{author}{Liu, X.}, \bibinfo{author}{Tao, F.}, \bibinfo{author}{Du, H.},
  \bibinfo{author}{Yu, W.}, \bibinfo{author}{Xu, K.}, \bibinfo{year}{2020}.
\newblock \bibinfo{title}{Learning {Nonlinear} {Constitutive} {Laws} {Using}
  {Neural} {Network} {Models} {Based} on {Indirectly} {Measurable} {Data}}.
\newblock \bibinfo{journal}{Journal of Applied Mechanics} \bibinfo{volume}{87},
  \bibinfo{pages}{081003}.
\newblock \URLprefix
  \url{https://asmedigitalcollection.asme.org/appliedmechanics/article/doi/10.1115/1.4047036/1083320/Learning-Nonlinear-Constitutive-Laws-Using-Neural},
  \DOIprefix\doi{10.1115/1.4047036}.
\bibitem[{Liu et~al.(2021)Liu, Tian, Tao, Du and Yu}]{liu_machine_2021}
\bibinfo{author}{Liu, X.}, \bibinfo{author}{Tian, S.}, \bibinfo{author}{Tao,
  F.}, \bibinfo{author}{Du, H.}, \bibinfo{author}{Yu, W.},
  \bibinfo{year}{2021}.
\newblock \bibinfo{title}{Machine learning-assisted modeling of composite
  materials and structures: a review}, in: \bibinfo{booktitle}{{AIAA} {Scitech}
  2021 {Forum}}, \bibinfo{publisher}{American Institute of Aeronautics and
  Astronautics}, \bibinfo{address}{VIRTUAL EVENT}.
\newblock \URLprefix \url{https://arc.aiaa.org/doi/10.2514/6.2021-2023},
  \DOIprefix\doi{10.2514/6.2021-2023}.
\bibitem[{Man and Furukawa(2011)}]{man_neural_2011}
\bibinfo{author}{Man, H.}, \bibinfo{author}{Furukawa, T.},
  \bibinfo{year}{2011}.
\newblock \bibinfo{title}{Neural network constitutive modelling for non-linear
  characterization of anisotropic materials}.
\newblock \bibinfo{journal}{International Journal for Numerical Methods in
  Engineering} \bibinfo{volume}{85}, \bibinfo{pages}{939--957}.
\newblock \URLprefix
  \url{https://onlinelibrary.wiley.com/doi/10.1002/nme.2999},
  \DOIprefix\doi{10.1002/nme.2999}.
\bibitem[{Marek et~al.(2019)Marek, Davis, Rossi and
  Pierron}]{marek_extension_2019}
\bibinfo{author}{Marek, A.}, \bibinfo{author}{Davis, F.M.},
  \bibinfo{author}{Rossi, M.}, \bibinfo{author}{Pierron, F.},
  \bibinfo{year}{2019}.
\newblock \bibinfo{title}{Extension of the sensitivity-based virtual fields to
  large deformation anisotropic plasticity}.
\newblock \bibinfo{journal}{International Journal of Material Forming}
  \bibinfo{volume}{12}, \bibinfo{pages}{457--476}.
\newblock \URLprefix \url{http://link.springer.com/10.1007/s12289-018-1428-1},
  \DOIprefix\doi{10.1007/s12289-018-1428-1}.
\bibitem[{Masi and Stefanou(2022)}]{masi_evolution_2022}
\bibinfo{author}{Masi, F.}, \bibinfo{author}{Stefanou, I.},
  \bibinfo{year}{2022}.
\newblock \bibinfo{title}{Evolution {TANN} and the discovery of the internal
  variables and evolution equations in solid mechanics}.
\newblock \URLprefix \url{http://arxiv.org/abs/2209.13269}.
  \bibinfo{note}{arXiv:2209.13269 [cond-mat]}.
\bibitem[{Masi et~al.(2021)Masi, Stefanou, Vannucci and
  Maffi-Berthier}]{masi_thermodynamics-based_2021}
\bibinfo{author}{Masi, F.}, \bibinfo{author}{Stefanou, I.},
  \bibinfo{author}{Vannucci, P.}, \bibinfo{author}{Maffi-Berthier, V.},
  \bibinfo{year}{2021}.
\newblock \bibinfo{title}{Thermodynamics-based {Artificial} {Neural} {Networks}
  for constitutive modeling}.
\newblock \bibinfo{journal}{Journal of the Mechanics and Physics of Solids}
  \bibinfo{volume}{147}, \bibinfo{pages}{104277}.
\newblock \URLprefix
  \url{https://linkinghub.elsevier.com/retrieve/pii/S0022509620304841},
  \DOIprefix\doi{10.1016/j.jmps.2020.104277}.
\bibitem[{Michel and Suquet(2016)}]{michel_model-reduction_2016}
\bibinfo{author}{Michel, J.C.}, \bibinfo{author}{Suquet, P.},
  \bibinfo{year}{2016}.
\newblock \bibinfo{title}{A model-reduction approach in micromechanics of
  materials preserving the variational structure of constitutive relations}.
\newblock \bibinfo{journal}{Journal of the Mechanics and Physics of Solids}
  \bibinfo{volume}{90}, \bibinfo{pages}{254--285}.
\newblock \URLprefix
  \url{https://linkinghub.elsevier.com/retrieve/pii/S0022509616300928},
  \DOIprefix\doi{10.1016/j.jmps.2016.02.005}.
\bibitem[{Miehe(2011)}]{miehe_multi-field_2011}
\bibinfo{author}{Miehe, C.}, \bibinfo{year}{2011}.
\newblock \bibinfo{title}{A multi-field incremental variational framework for
  gradient-extended standard dissipative solids}.
\newblock \bibinfo{journal}{J. Mech. Phys. Solids} , \bibinfo{pages}{26}.
\bibitem[{Mielke(2011)}]{mielke_formulation_2011}
\bibinfo{author}{Mielke, A.}, \bibinfo{year}{2011}.
\newblock \bibinfo{title}{Formulation of thermoelastic dissipative material
  behavior using {GENERIC}}.
\newblock \bibinfo{journal}{Continuum Mechanics and Thermodynamics}
  \bibinfo{volume}{23}, \bibinfo{pages}{233--256}.
\newblock \URLprefix \url{http://link.springer.com/10.1007/s00161-010-0179-0},
  \DOIprefix\doi{10.1007/s00161-010-0179-0}.
\bibitem[{Miled(2011)}]{miled_coupled_2011-1}
\bibinfo{author}{Miled, B.}, \bibinfo{year}{2011}.
\newblock \bibinfo{title}{Coupled viscoelastic-viscoplastic modeling of
  homogeneous and reinforced thermoplastic polymers}.
\newblock Ph.D. thesis.
\bibitem[{Miled et~al.(2011)Miled, Doghri and Delannay}]{miled_coupled_2011}
\bibinfo{author}{Miled, B.}, \bibinfo{author}{Doghri, I.},
  \bibinfo{author}{Delannay, L.}, \bibinfo{year}{2011}.
\newblock \bibinfo{title}{Coupled viscoelastic–viscoplastic modeling of
  homogeneous and isotropic polymers: {Numerical} algorithm and analytical
  solutions}.
\newblock \bibinfo{journal}{Computer Methods in Applied Mechanics and
  Engineering} \bibinfo{volume}{200}, \bibinfo{pages}{3381--3394}.
\newblock \URLprefix
  \url{https://linkinghub.elsevier.com/retrieve/pii/S0045782511002702},
  \DOIprefix\doi{10.1016/j.cma.2011.08.015}.
\bibitem[{Montáns et~al.(2019)Montáns, Chinesta, Gómez-Bombarelli and
  Kutz}]{montans_data-driven_2019}
\bibinfo{author}{Montáns, F.J.}, \bibinfo{author}{Chinesta, F.},
  \bibinfo{author}{Gómez-Bombarelli, R.}, \bibinfo{author}{Kutz, J.N.},
  \bibinfo{year}{2019}.
\newblock \bibinfo{title}{Data-driven modeling and learning in science and
  engineering}.
\newblock \bibinfo{journal}{Comptes Rendus Mécanique} \bibinfo{volume}{347},
  \bibinfo{pages}{845--855}.
\newblock \URLprefix
  \url{https://linkinghub.elsevier.com/retrieve/pii/S1631072119301809},
  \DOIprefix\doi{10.1016/j.crme.2019.11.009}.
\bibitem[{Neggers et~al.(2018)Neggers, Allix, Hild and Roux}]{neggers_big_2018}
\bibinfo{author}{Neggers, J.}, \bibinfo{author}{Allix, O.},
  \bibinfo{author}{Hild, F.}, \bibinfo{author}{Roux, S.}, \bibinfo{year}{2018}.
\newblock \bibinfo{title}{Big {Data} in {Experimental} {Mechanics} and {Model}
  {Order} {Reduction}: {Today}’s {Challenges} and {Tomorrow}’s
  {Opportunities}}.
\newblock \bibinfo{journal}{Archives of Computational Methods in Engineering}
  \bibinfo{volume}{25}, \bibinfo{pages}{143--164}.
\newblock \URLprefix \url{http://link.springer.com/10.1007/s11831-017-9234-3},
  \DOIprefix\doi{10.1007/s11831-017-9234-3}.
\bibitem[{Neto et~al.(2008)Neto, Peric and Owen}]{neto_computational_2008}
\bibinfo{author}{Neto, E.d.S.}, \bibinfo{author}{Peric, D.},
  \bibinfo{author}{Owen, D.}, \bibinfo{year}{2008}.
\newblock \bibinfo{title}{Computational methods for plasticity}.
\newblock \bibinfo{publisher}{John Wiley \& Sons}.
\bibitem[{Nguyen(2010)}]{nguyen_standard_2010}
\bibinfo{author}{Nguyen, Q.S.}, \bibinfo{year}{2010}.
\newblock \bibinfo{title}{On standard dissipative gradient models}.
\newblock \bibinfo{journal}{Annals of Solid and Structural Mechanics}
  \bibinfo{volume}{1}, \bibinfo{pages}{79--86}.
\newblock \URLprefix \url{http://link.springer.com/10.1007/s12356-010-0006-0},
  \DOIprefix\doi{10.1007/s12356-010-0006-0}.
\bibitem[{Onsager(1931a)}]{onsager_reciprocal_1931-1}
\bibinfo{author}{Onsager, L.}, \bibinfo{year}{1931}a.
\newblock \bibinfo{title}{Reciprocal {Relations} in {Irreversible} {Processes}.
  {I}.}
\newblock \bibinfo{journal}{Physical Review} \bibinfo{volume}{37},
  \bibinfo{pages}{405--426}.
\newblock \URLprefix \url{https://link.aps.org/doi/10.1103/PhysRev.37.405},
  \DOIprefix\doi{10.1103/PhysRev.37.405}.
\bibitem[{Onsager(1931b)}]{onsager_reciprocal_1931}
\bibinfo{author}{Onsager, L.}, \bibinfo{year}{1931}b.
\newblock \bibinfo{title}{Reciprocal {Relations} in {Irreversible} {Processes}.
  {II}.}
\newblock \bibinfo{journal}{Physical Review} \bibinfo{volume}{38},
  \bibinfo{pages}{2265--2279}.
\newblock \URLprefix \url{https://link.aps.org/doi/10.1103/PhysRev.38.2265},
  \DOIprefix\doi{10.1103/PhysRev.38.2265}.
\bibitem[{Peng et~al.(2021)Peng, Alber, Buganza~Tepole, Cannon, De,
  Dura-Bernal, Garikipati, Karniadakis, Lytton, Perdikaris, Petzold and
  Kuhl}]{peng_multiscale_2021}
\bibinfo{author}{Peng, G.C.Y.}, \bibinfo{author}{Alber, M.},
  \bibinfo{author}{Buganza~Tepole, A.}, \bibinfo{author}{Cannon, W.R.},
  \bibinfo{author}{De, S.}, \bibinfo{author}{Dura-Bernal, S.},
  \bibinfo{author}{Garikipati, K.}, \bibinfo{author}{Karniadakis, G.},
  \bibinfo{author}{Lytton, W.W.}, \bibinfo{author}{Perdikaris, P.},
  \bibinfo{author}{Petzold, L.}, \bibinfo{author}{Kuhl, E.},
  \bibinfo{year}{2021}.
\newblock \bibinfo{title}{Multiscale {Modeling} {Meets} {Machine} {Learning}:
  {What} {Can} {We} {Learn}?}
\newblock \bibinfo{journal}{Archives of Computational Methods in Engineering}
  \bibinfo{volume}{28}, \bibinfo{pages}{1017--1037}.
\newblock \URLprefix
  \url{https://link.springer.com/10.1007/s11831-020-09405-5},
  \DOIprefix\doi{10.1007/s11831-020-09405-5}.
\bibitem[{Pierron et~al.(2010)Pierron, Avril and
  The~Tran}]{pierron_extension_2010}
\bibinfo{author}{Pierron, F.}, \bibinfo{author}{Avril, S.},
  \bibinfo{author}{The~Tran, V.}, \bibinfo{year}{2010}.
\newblock \bibinfo{title}{Extension of the virtual fields method to
  elasto-plastic material identification with cyclic loads and kinematic
  hardening}.
\newblock \bibinfo{journal}{International Journal of Solids and Structures}
  \bibinfo{volume}{47}, \bibinfo{pages}{2993--3010}.
\bibitem[{Pierron and Grédiac(2012)}]{pierron_virtual_2012}
\bibinfo{author}{Pierron, F.}, \bibinfo{author}{Grédiac, M.},
  \bibinfo{year}{2012}.
\newblock \bibinfo{title}{The {Virtual} {Fields} {Method}}.
\newblock \bibinfo{publisher}{Springer New York}, \bibinfo{address}{New York,
  NY}.
\newblock \URLprefix \url{http://link.springer.com/10.1007/978-1-4614-1824-5},
  \DOIprefix\doi{10.1007/978-1-4614-1824-5}.
\bibitem[{Rockafellar(2015)}]{rockafellar_convex_2015}
\bibinfo{author}{Rockafellar, R.T.}, \bibinfo{year}{2015}.
\newblock \bibinfo{title}{Convex {Analysis}}.
\newblock \bibinfo{publisher}{Princeton University Press}.
\newblock \URLprefix \url{https://doi.org/10.1515/9781400873173},
  \DOIprefix\doi{10.1515/9781400873173}.
\bibitem[{Rothe and Hartmann(2015)}]{rothe_automatic_2015}
\bibinfo{author}{Rothe, S.}, \bibinfo{author}{Hartmann, S.},
  \bibinfo{year}{2015}.
\newblock \bibinfo{title}{Automatic differentiation for stress and consistent
  tangent computation}.
\newblock \bibinfo{journal}{Archive of Applied Mechanics} \bibinfo{volume}{85},
  \bibinfo{pages}{1103--1125}.
\newblock \URLprefix \url{http://link.springer.com/10.1007/s00419-014-0939-6},
  \DOIprefix\doi{10.1007/s00419-014-0939-6}.
\bibitem[{Roux and Hild(2020)}]{roux_optimal_2020}
\bibinfo{author}{Roux, S.}, \bibinfo{author}{Hild, F.}, \bibinfo{year}{2020}.
\newblock \bibinfo{title}{Optimal procedure for the identification of
  constitutive parameters from experimentally measured displacement fields}.
\newblock \bibinfo{journal}{International Journal of Solids and Structures}
  \bibinfo{volume}{184}, \bibinfo{pages}{14--23}.
\newblock \URLprefix
  \url{https://linkinghub.elsevier.com/retrieve/pii/S0020768318304542},
  \DOIprefix\doi{10.1016/j.ijsolstr.2018.11.008}.
\bibitem[{Savitzky and Golay(1964)}]{savitzky_smoothing_1964}
\bibinfo{author}{Savitzky, A.}, \bibinfo{author}{Golay, M.J.E.},
  \bibinfo{year}{1964}.
\newblock \bibinfo{title}{Smoothing and {Differentiation} of {Data} by
  {Simplified} {Least} {Squares} {Procedures}.}
\newblock \bibinfo{journal}{Analytical Chemistry} \bibinfo{volume}{36},
  \bibinfo{pages}{1627--1639}.
\newblock \URLprefix \url{https://pubs.acs.org/doi/abs/10.1021/ac60214a047},
  \DOIprefix\doi{10.1021/ac60214a047}.
\bibitem[{Schmidt and Lipson(2009)}]{schmidt_distilling_2009}
\bibinfo{author}{Schmidt, M.}, \bibinfo{author}{Lipson, H.},
  \bibinfo{year}{2009}.
\newblock \bibinfo{title}{Distilling {Free}-{Form} {Natural} {Laws} from
  {Experimental} {Data}}.
\newblock \bibinfo{journal}{Science} \bibinfo{volume}{324},
  \bibinfo{pages}{81--85}.
\newblock \URLprefix
  \url{https://www.sciencemag.org/lookup/doi/10.1126/science.1165893},
  \DOIprefix\doi{10.1126/science.1165893}.
\bibitem[{Simo and Hughes(1998)}]{simo_computational_1998}
\bibinfo{author}{Simo, J.C.}, \bibinfo{author}{Hughes, T.J.R.},
  \bibinfo{year}{1998}.
\newblock \bibinfo{title}{Computational inelasticity}.
\newblock Number \bibinfo{number}{v. 7} in \bibinfo{series}{Interdisciplinary
  applied mathematics}, \bibinfo{publisher}{Springer}, \bibinfo{address}{New
  York}.
\bibitem[{Stainier et~al.(2019)Stainier, Leygue and
  Ortiz}]{stainier_model-free_2019}
\bibinfo{author}{Stainier, L.}, \bibinfo{author}{Leygue, A.},
  \bibinfo{author}{Ortiz, M.}, \bibinfo{year}{2019}.
\newblock \bibinfo{title}{Model-free data-driven methods in mechanics: material
  data identification and solvers}.
\newblock \bibinfo{journal}{Computational Mechanics} \bibinfo{volume}{64},
  \bibinfo{pages}{381--393}.
\newblock \URLprefix \url{http://link.springer.com/10.1007/s00466-019-01731-1},
  \DOIprefix\doi{10.1007/s00466-019-01731-1}.
\bibitem[{Steinmann and Runesson(2021)}]{steinmann_catalogue_2021}
\bibinfo{author}{Steinmann, P.}, \bibinfo{author}{Runesson, K.},
  \bibinfo{year}{2021}.
\newblock \bibinfo{title}{The {Catalogue} of {Computational} {Material}
  {Models}: {Basic} {Geometrically} {Linear} {Models} in {1D}}.
\newblock \bibinfo{publisher}{Springer International Publishing},
  \bibinfo{address}{Cham}.
\newblock \URLprefix \url{http://link.springer.com/10.1007/978-3-030-63684-5},
  \DOIprefix\doi{10.1007/978-3-030-63684-5}.
\bibitem[{Sun and Zhu(2013)}]{sun_serial_2013}
\bibinfo{author}{Sun, L.}, \bibinfo{author}{Zhu, Y.}, \bibinfo{year}{2013}.
\newblock \bibinfo{title}{A serial two-stage viscoelastic–viscoplastic
  constitutive model with thermodynamical consistency for characterizing
  time-dependent deformation behavior of asphalt concrete mixtures}.
\newblock \bibinfo{journal}{Construction and Building Materials}
  \bibinfo{volume}{40}, \bibinfo{pages}{584--595}.
\newblock \URLprefix
  \url{https://linkinghub.elsevier.com/retrieve/pii/S0950061812007489},
  \DOIprefix\doi{10.1016/j.conbuildmat.2012.10.004}.
\bibitem[{Sussman and Bathe(2009)}]{sussman_model_2009}
\bibinfo{author}{Sussman, T.}, \bibinfo{author}{Bathe, K.J.},
  \bibinfo{year}{2009}.
\newblock \bibinfo{title}{A model of incompressible isotropic hyperelastic
  material behavior using spline interpolations of tension-compression test
  data}.
\newblock \bibinfo{journal}{Communications in Numerical Methods in Engineering}
  \bibinfo{volume}{25}, \bibinfo{pages}{53--63}.
\newblock \URLprefix \url{http://doi.wiley.com/10.1002/cnm.1105},
  \DOIprefix\doi{10.1002/cnm.1105}.
\bibitem[{Tac et~al.(2022)Tac, Sahli~Costabal and
  Tepole}]{tac_data-driven_2022}
\bibinfo{author}{Tac, V.}, \bibinfo{author}{Sahli~Costabal, F.},
  \bibinfo{author}{Tepole, A.B.}, \bibinfo{year}{2022}.
\newblock \bibinfo{title}{Data-driven tissue mechanics with polyconvex neural
  ordinary differential equations}.
\newblock \bibinfo{journal}{Computer Methods in Applied Mechanics and
  Engineering} \bibinfo{volume}{398}, \bibinfo{pages}{115248}.
\newblock \URLprefix
  \url{https://linkinghub.elsevier.com/retrieve/pii/S0045782522003838},
  \DOIprefix\doi{10.1016/j.cma.2022.115248}.
\bibitem[{Thakolkaran et~al.(2022)Thakolkaran, Joshi, Zheng, Flaschel,
  De~Lorenzis and Kumar}]{thakolkaran_nn-euclid_2022}
\bibinfo{author}{Thakolkaran, P.}, \bibinfo{author}{Joshi, A.},
  \bibinfo{author}{Zheng, Y.}, \bibinfo{author}{Flaschel, M.},
  \bibinfo{author}{De~Lorenzis, L.}, \bibinfo{author}{Kumar, S.},
  \bibinfo{year}{2022}.
\newblock \bibinfo{title}{{NN}-{EUCLID}: {Deep}-learning hyperelasticity
  without stress data}.
\newblock \bibinfo{journal}{Journal of the Mechanics and Physics of Solids}
  \bibinfo{volume}{169}, \bibinfo{pages}{105076}.
\newblock \URLprefix
  \url{https://linkinghub.elsevier.com/retrieve/pii/S0022509622002538},
  \DOIprefix\doi{10.1016/j.jmps.2022.105076}.
\bibitem[{Tibshirani(1996)}]{tibshirani_regression_1996}
\bibinfo{author}{Tibshirani, R.}, \bibinfo{year}{1996}.
\newblock \bibinfo{title}{Regression {Shrinkage} and {Selection} via the
  {Lasso}}.
\newblock \bibinfo{journal}{Journal of the Royal Statistical Society: Series B
  (Methodological)} \bibinfo{volume}{58}, \bibinfo{pages}{267--288}.
\newblock \URLprefix
  \url{http://doi.wiley.com/10.1111/j.2517-6161.1996.tb02080.x},
  \DOIprefix\doi{10.1111/j.2517-6161.1996.tb02080.x}.
\bibitem[{Touchette(2007)}]{touchette_legendre-fenchel_2007}
\bibinfo{author}{Touchette, H.}, \bibinfo{year}{2007}.
\newblock \bibinfo{title}{Legendre-{Fenchel} transforms in a nutshell} .
\bibitem[{Vlassis and Sun(2021)}]{vlassis_sobolev_2021}
\bibinfo{author}{Vlassis, N.N.}, \bibinfo{author}{Sun, W.},
  \bibinfo{year}{2021}.
\newblock \bibinfo{title}{Sobolev training of thermodynamic-informed neural
  networks for interpretable elasto-plasticity models with level set
  hardening}.
\newblock \bibinfo{journal}{Computer Methods in Applied Mechanics and
  Engineering} \bibinfo{volume}{377}, \bibinfo{pages}{113695}.
\newblock \URLprefix
  \url{https://linkinghub.elsevier.com/retrieve/pii/S0045782521000311},
  \DOIprefix\doi{10.1016/j.cma.2021.113695}.
\bibitem[{Wang et~al.(2021)Wang, Estrada, Arruda and
  Garikipati}]{wang_inference_2021}
\bibinfo{author}{Wang, Z.}, \bibinfo{author}{Estrada, J.},
  \bibinfo{author}{Arruda, E.}, \bibinfo{author}{Garikipati, K.},
  \bibinfo{year}{2021}.
\newblock \bibinfo{title}{Inference of deformation mechanisms and constitutive
  response of soft material surrogates of biological tissue by full-field
  characterization and data-driven variational system identification}.
\newblock \bibinfo{journal}{Journal of the Mechanics and Physics of Solids}
  \bibinfo{volume}{153}, \bibinfo{pages}{104474}.
\newblock \URLprefix
  \url{https://linkinghub.elsevier.com/retrieve/pii/S0022509621001459},
  \DOIprefix\doi{10.1016/j.jmps.2021.104474}.
\bibitem[{Yu et~al.(2021)Yu, Tian, E and Li}]{yu_onsagernet_2021}
\bibinfo{author}{Yu, H.}, \bibinfo{author}{Tian, X.}, \bibinfo{author}{E, W.},
  \bibinfo{author}{Li, Q.}, \bibinfo{year}{2021}.
\newblock \bibinfo{title}{{OnsagerNet}: {Learning} stable and interpretable
  dynamics using a generalized {Onsager} principle}.
\newblock \bibinfo{journal}{Physical Review Fluids} \bibinfo{volume}{6},
  \bibinfo{pages}{114402}.
\newblock \URLprefix
  \url{https://link.aps.org/doi/10.1103/PhysRevFluids.6.114402},
  \DOIprefix\doi{10.1103/PhysRevFluids.6.114402}.
\bibitem[{Zhang et~al.(2022)Zhang, Shin and Karniadakis}]{zhang_gfinns_2022}
\bibinfo{author}{Zhang, Z.}, \bibinfo{author}{Shin, Y.},
  \bibinfo{author}{Karniadakis, G.E.}, \bibinfo{year}{2022}.
\newblock \bibinfo{title}{{GFINNs}: {GENERIC} {Formalism} {Informed} {Neural}
  {Networks} for {Deterministic} and {Stochastic} {Dynamical} {Systems}}.
\newblock \bibinfo{journal}{Philosophical Transactions of the Royal Society A:
  Mathematical, Physical and Engineering Sciences} \bibinfo{volume}{380},
  \bibinfo{pages}{20210207}.
\newblock \URLprefix \url{http://arxiv.org/abs/2109.00092},
  \DOIprefix\doi{10.1098/rsta.2021.0207}. \bibinfo{note}{arXiv:2109.00092 [cs,
  math]}.
\bibitem[{Ziegler(1957)}]{ziegler_thermodynamik_1957}
\bibinfo{author}{Ziegler, H.}, \bibinfo{year}{1957}.
\newblock \bibinfo{title}{Thermodynamik und rheologische {Probleme}}.
\newblock \bibinfo{journal}{Ingenieur-Archiv} \bibinfo{volume}{25},
  \bibinfo{pages}{58--70}.
\newblock \URLprefix \url{http://link.springer.com/10.1007/BF00536645},
  \DOIprefix\doi{10.1007/BF00536645}.
\bibitem[{Ziegler(1958)}]{ziegler_attempt_1958}
\bibinfo{author}{Ziegler, H.}, \bibinfo{year}{1958}.
\newblock \bibinfo{title}{An attempt to generalize {Onsager}'s principle, and
  its significance for rheological problems}.
\newblock \bibinfo{journal}{ZAMP Zeitschrift für Angewandte Mathematik und
  Physik} \bibinfo{volume}{9}, \bibinfo{pages}{748--763}.
\newblock \URLprefix \url{http://link.springer.com/10.1007/BF02424793},
  \DOIprefix\doi{10.1007/BF02424793}.
\bibitem[{Ziegler(1972)}]{ziegler_systems_1972}
\bibinfo{author}{Ziegler, H.}, \bibinfo{year}{1972}.
\newblock \bibinfo{title}{Systems with internal parameters obeying the
  orthogonality condition}.
\newblock \bibinfo{journal}{Zeitschrift für angewandte Mathematik und Physik
  ZAMP} \bibinfo{volume}{23}, \bibinfo{pages}{553--566}.
\newblock \URLprefix \url{http://link.springer.com/10.1007/BF01593978},
  \DOIprefix\doi{10.1007/BF01593978}.

\end{thebibliography}

\end{document}